# Hybrid BDI-POMDP Framework for Multiagent Teaming

**Ranjit Nair**                                                    RANJIT.NAIR@HONEYWELL.COM
*Automation and Control Solutions*
*Honeywell Laboratories, Minneapolis, MN 55416*
**Milind Tambe**                                                            TAMBE@USC.EDU
*Department of Computer Science*
*University of Southern California, Los Angeles, CA 90089*

## Abstract

Many current large-scale multiagent team implementations can be characterized as following the "belief-desire-intention" (BDI) paradigm, with explicit representation of team plans. Despite their promise, current BDI team approaches lack tools for quantitative performance analysis under uncertainty. Distributed partially observable Markov decision problems (POMDPs) are well suited for such analysis, but the complexity of finding optimal policies in such models is highly intractable. The key contribution of this article is a hybrid BDI-POMDP approach, where BDI team plans are exploited to improve POMDP tractability and POMDP analysis improves BDI team plan performance.

Concretely, we focus on role allocation, a fundamental problem in BDI teams: which agents to allocate to the different roles in the team. The article provides three key contributions. First, we describe a role allocation technique that takes into account future uncertainties in the domain; prior work in multiagent role allocation has failed to address such uncertainties. To that end, we introduce RMTDP (Role-based Markov Team Decision Problem), a new distributed POMDP model for analysis of role allocations. Our technique gains in tractability by significantly curtailing RMTDP policy search; in particular, BDI team plans provide incomplete RMTDP policies, and the RMTDP policy search fills the gaps in such incomplete policies by searching for the best role allocation. Our second key contribution is a novel decomposition technique to further improve RMTDP policy search efficiency. Even though limited to searching role allocations, there are still combinatorially many role allocations, and evaluating each in RMTDP to identify the best is extremely difficult. Our decomposition technique exploits the structure in the BDI team plans to significantly prune the search space of role allocations. Our third key contribution is a significantly faster policy evaluation algorithm suited for our BDI-POMDP hybrid approach. Finally, we also present experimental results from two domains: mission rehearsal simulation and RoboCupRescue disaster rescue simulation.

## 1. Introduction

Teamwork, whether among software agents, or robots (and people) is a critical capability in a large number of multiagent domains ranging from mission rehearsal simulations, to RoboCup soccer and disaster rescue, to personal assistant teams. Already a large number of multiagent teams have been developed for a range of domains (Pynadath & Tambe, 2003; Yen, Yin, Ioerger, Miller, Xu, & Volz, 2001; Stone & Veloso, 1999; Jennings, 1995; Grosz, Hunsberger, & Kraus, 1999; Decker & Lesser, 1993; Tambe, Pynadath, & Chauvat, 2000; da Silva & Demazeau, 2002). These existing practical approaches can be characterized as situated within the general "belief-desire-intention" (BDI) approach, a paradigm for designing multiagent systems, made increasingly popular due to programming frameworks (Tambe et al., 2000; Decker & Lesser, 1993; Tidhar, 1993b) that facilitate the design of large-scale teams. Within this approach, inspired explicitly or implicitly by BDI logics, agents explicitly represent and reason with their team goals and plans (Wooldridge, 2002).

This article focuses on analysis of such BDI teams, to provide feedback to aid human developers and possibly to agents participating in a team, on how the team performance in complex dynamic domains can be improved. In particular, it focuses on the critical challenge of role allocation in building teams (Tidhar, Rao, & Sonenberg, 1996; Hunsberger & Grosz, 2000), i.e. which agents to allocate to the various roles in the team. For instance, in mission rehearsal simulations (Tambe et al., 2000), we need to select the numbers and types of helicopter agents to allocate to different roles in the team. Similarly, in disaster rescue (Kitano, Tadokoro, Noda, Matsubara, Takahashi, Shinjoh, & Shimada, 1999), role allocation refers to allocating fire engines and ambulances to fires and it can greatly impact team performance. In both these and other such domains, the performance of the team is linked to important metrics such as loss of human life and property and thus it is critical to analyze team performance and suggest improvements.

While BDI frameworks facilitate human design of large scale teams, the key difficulty in analyzing role allocation in these teams is due to the uncertainty that arises in complex domains. For example, actions may fail and the world state may be only *partially observable* to the agents owing to physical properties of the environment or imperfect sensing. Role allocation demands such future uncertainties be taken into account, e.g. the fact that an agent may fail during execution and may or may not be replaced by another must be taken into account when determining the role allocation. Yet most current role allocation algorithms do not address such uncertainty (see Section 7.4). Indeed, such uncertainty requires quantitative comparison of different role allocations. However, tools for such quantitative evaluations of BDI teams are currently absent. Thus, given these uncertainties, we may be required to experimentally recreate a large number of possible scenarios (in a real domain or in simulations) to evaluate and compare different role allocations.

Fortunately, the emergence of distributed Partially Observable Markov Decision Problems (POMDPs) provides models (Bernstein, Zilberstein, & Immerman, 2000; Boutilier, 1996; Pynadath & Tambe, 2002; Xuan, Lesser, & Zilberstein, 2001) that can be used for quantitative analysis of agent teams in uncertain domains. Distributed POMDPs represent a class of formal models that are powerful enough to express the uncertainty in these dynamic domains arising as a result of *non-determinism* and *partial observability* and in principle, can be used to generate and evaluate complete policies for the multiagent team. However, there are two shortcomings in these models that prevents their application in the analysis of role allocation. First, previous work on analysis has focused on communication (Pynadath & Tambe, 2002; Xuan et al., 2001), rather than role allocation or any other coordination decisions. Second, as shown by Bernstein *et al.* (2000), the problem of deriving the optimal policy is generally computationally intractable (the corresponding decision problem is NEXP-complete). Thus, applying optimal policies for analysis is highly intractable.

To address the first difficulty, we derive RMTDP (Role-based Multiagent Team Decision Problem), a distributed POMDP framework for quantitatively analyzing role allocations. Using this framework, we show that, in general, the problem of finding the optimal role allocation policy is computationally intractable (the corresponding decision problem is still NEXP-complete). This shows that improving the tractability of analysis techniques for role allocation is a critically important issue.

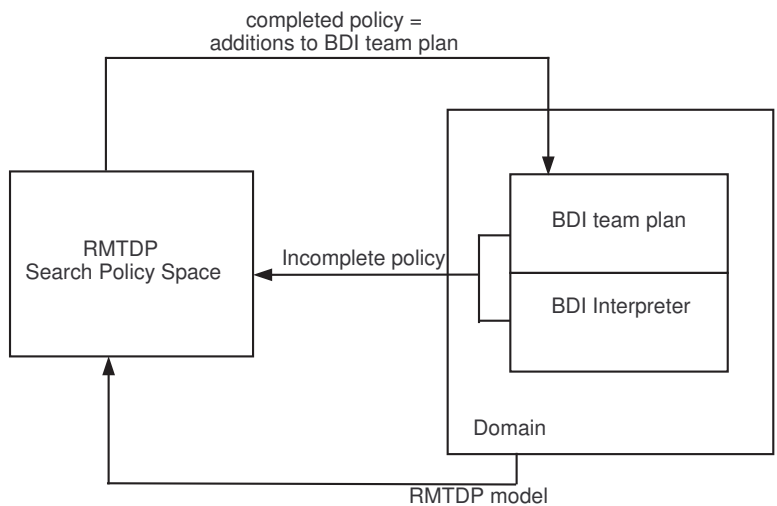

Figure 1: Integration of BDI and POMDP.

Therefore, in order to make the quantitative analysis of multiagent teams using RMTDP more tractable, our second contribution provides a hybrid BDI-POMDP approach that combines the native strengths of the BDI and POMDP approaches, i.e., the ability in BDI frameworks to encode large-scale team plans and the POMDP ability to quantitatively evaluate such plans. This hybrid approach is based on three key interactions that improve the tractability of RMTDP and the optimality of BDI agent teams. The first interaction is shown in Figure 1. In particular, suppose we wish to analyze a BDI agent team (each agent consisting of a BDI team plan and a domain independent interpreter that helps coordinate such plans) acting in a domain. Then as shown in Figure 1, we model the domain via an RMTDP, and rely on the BDI team plan and interpreter for providing an incomplete policy for this RMTDP. The RMTDP model evaluates different completions of this incomplete policy and provides an optimally completed policy as feedback to the BDI system. Thus, the RMTDP fills in the gaps in an incompletely specified BDI team plan optimally. Here the gaps we concentrate on are the role allocations, but the method can be applied to other key coordination decisions. By restricting the optimization to only role allocation decisions and fixing the policy at all other points, we are able to come up with a restricted policy space. We then use RMTDPs to effectively search this restricted space in order to find the optimal role allocation.

While the restricted policy search is one key positive interaction in our hybrid approach, the second interaction consists of a more efficient policy representation used for converting a BDI team plan and interpreter into a corresponding policy (see Figure 1) and a new algorithm for policy evaluation. In general, each agent's policy in a distributed POMDP is indexed by its observation history (Bernstein et al., 2000; Pynadath & Tambe, 2002).

However, in a BDI system, each agent performs its action selection based on its set of privately held beliefs which is obtained from the agent's observations after applying a belief revision function. In order to evaluate the team's performance, it is sufficient in RMTDP to index the agents' policies by their *belief state* (represented here by their privately held beliefs) instead of their observation histories. This shift in representation results in considerable savings in the amount of time needed to evaluate a policy and in the space required to represent a policy.

The third key interaction in our hybrid approach further exploits BDI team plan structure for increasing the efficiency of our RMTDP-based analysis. Even though RMTDP policy space is restricted to filling in gaps in incomplete policies, many policies may result given the large number of possible role allocations. Thus enumerating and evaluating each possible policy for a given domain is difficult. Instead, we provide a branch-and-bound algorithm that exploits task decomposition among sub-teams of a team to significantly prune the search space and provide a correctness proof and worst-case analysis of this algorithm.

In order to empirically validate our approach, we have applied RMTDP for allocation in BDI teams in two concrete domains: mission rehearsal simulations (Tambe et al., 2000) and RoboCupRescue (Kitano et al., 1999). We first present the (significant) speed-up gained by our three interactions mentioned above. Next, in both domains, we compared the role allocations found by our approach with state-of-the-art techniques that allocate roles without uncertainty reasoning. This comparison shows the importance of reasoning about uncertainty when determining the role allocation for complex multiagent domains. In the RoboCupRescue domain, we also compared the allocations found with allocations chosen by humans in the actual RoboCupRescue simulation environment. The results showed that the role allocation technique presented in this article is capable of performing at human expert levels in the RoboCupRescue domain.

The article is organized as follows: Section 2 presents background and motivation. In Section 3, we introduce the RMTDP model and present key complexity results. Section 4 explains how a BDI team plan can be evaluated using RMTDP. Section 5 describes the analysis methodology for finding the optimal role allocation, while Section 6 presents an empirical evaluation of this methodology. In Section 7, we present related work and in Section 8, we list our conclusions.

## 2. Background

This section first describes the two domains that we consider in this article: an abstract mission rehearsal domain (Tambe et al., 2000) and the RoboCupRescue domain (Kitano et al., 1999). Each domain requires us to allocate roles to agents in a team. Next, team-oriented programming (TOP), a framework for describing team plans is described in the context of these two domains. While we focus on TOP, as discussed further in Section 7.1, our techniques would be applicable in other frameworks for tasking teams (Stone & Veloso, 1999; Decker & Lesser, 1993).

### 2.1 Domains

The first domain that we consider is based on mission rehearsal simulations (Tambe et al., 2000). For expository purposes, this has been intentionally simplified. The scenario is as follows: A helicopter team is executing a mission of transporting valuable cargo from point X to point Y through enemy terrain (see Figure 2). There are three paths from X to Y of different lengths and different risk due to enemy fire. One or more scouting sub-teams must be sent out (one for each path from X to Y), and the larger the size of a scouting sub-team the safer it is. When scouts clear up any one path from X to Y, the transports can then move more safely along that path. However, the scouts may fail along a path, and may need to be replaced by a transport at the cost of not transporting cargo. Owing to partial observability, the transports may not receive an observation that a scout has failed or that a route has been cleared. We wish to transport the most amount of cargo in the quickest possible manner within the mission deadline.

The key role allocation decision here is given a fixed number of helicopters, how should they be allocated to scouting and transport roles? Allocating more scouts means that the scouting task is more likely to succeed, but there will be fewer helicopters left that can be used to transport the cargo and consequently less reward. However, allocating too few scouts could result in the mission failing altogether. Also, in allocating the scouts, which routes should the scouts be sent on? The shortest route would be preferable but it is more risky. Sending all the scouts on the same route decreases the likelihood of failure of an individual scout; however, it might be more beneficial to send them on different routes, e.g. some scouts on a risky but short route and others on a safe but longer route.

Thus there are many role allocations to consider. Evaluating each is difficult because role allocation must look-ahead to consider future implications of uncertainty, e.g. scout helicopters can fail during scouting and may need to be replaced by a transport. Furthermore, failure or success of a scout may not be visible to the transport helicopters and hence a transport may not replace a scout or transports may never fly to the destination.

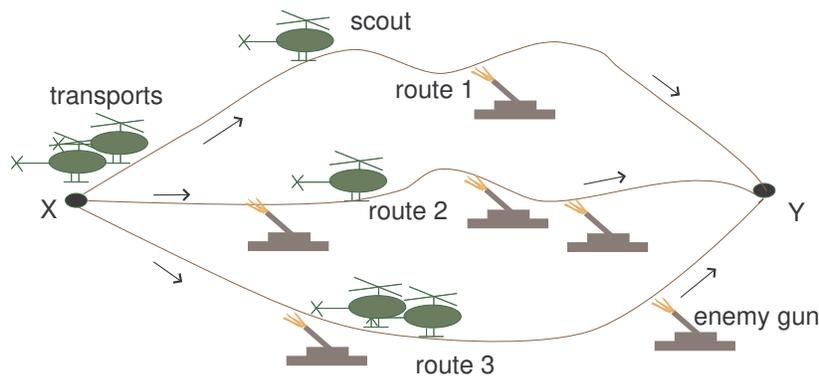

Figure 2: Mission rehearsal domain.

The second example scenario (see Figure 3), set up in the RoboCupRescue disaster simulation environment (Kitano et al., 1999), consists of five fire engines at three different fire stations (two each at stations 1 & 3 and the last at station 2) and five ambulances stationed at the ambulance center. Two fires (in top left and bottom right corners of the map) start that need to be extinguished by the fire engines. After a fire is extinguished, ambulance agents need to save the surviving civilians. The number of civilians at each location is not known ahead of time, although the total number of civilians in known. As time passes, there is a high likelihood that the health of civilians will deteriorate and fires will increase in intensity. Yet the agents need to rescue as many civilians as possible with minimal damage to the buildings. The first part of the goal in this scenario is therefore to first determine which fire engines to assign to each fire. Once the fire engines have gathered information about the number of civilians at each fire, this is transmitted to the ambulances. The next part of the goal is then to allocate the ambulances to a particular fire to rescue the civilians trapped there. However, ambulances cannot rescue civilians until fires are fully extinguished. Here, partial observability (each agent can only view objects within its visual range), and uncertainty related to fire intensity, as well as location of civilians and their health add significantly to the difficulty.

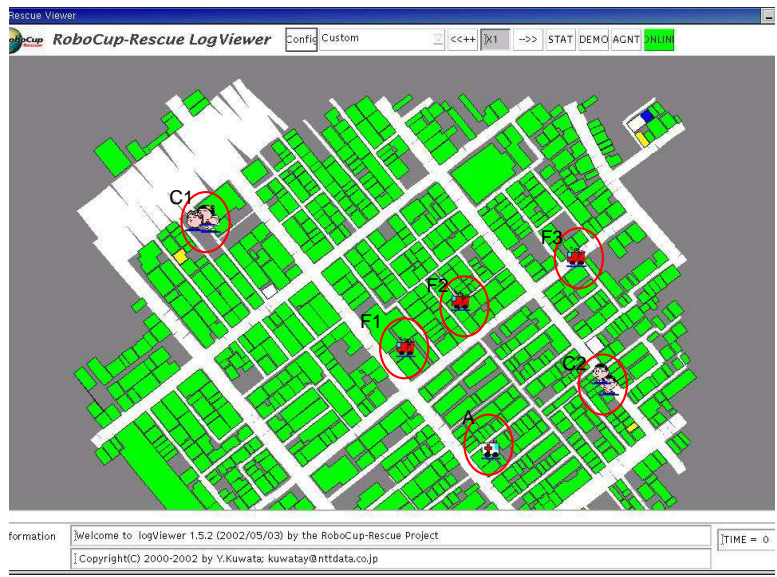

Figure 3: RoboCupRescue Scenario: C1 and C2 denote the two fire locations, F1, F2 and F3 denote fire stations 1, 2 and 3 respectively and A denotes the ambulance center.

## 2.2 Team-Oriented Programming

The aim of the team-oriented programming (TOP) (Pynadath & Tambe, 2003; Tambe et al., 2000; Tidhar, 1993b) framework is to provide human developers (or automated symbolic planners) with a useful abstraction for tasking teams. For domains such as those described in Section 2.1, it consists of three key aspects of a team: (i) a team organization hierarchy consisting of roles; (ii) a team (reactive) plan hierarchy; and (iii) an assignment of roles to sub-plans in the plan hierarchy. The developer need not specify low-level coordination details. Instead, the TOP interpreter (the underlying coordination infrastructure) automatically enables agents to decide when and with whom to communicate and how to reallocate roles upon failure. The TOP abstraction enables humans to rapidly provide team plans for large-scale teams, but unfortunately, only a qualitative assessment of team performance is feasible. Thus, a key TOP weakness is the inability to quantitatively evaluate and optimize team performance, e.g., in allocating roles to agents only a qualitative matching of capabilities may be feasible. As discussed later, our hybrid BDI-POMDP model addresses this weakness by providing techniques for quantitative evaluation.

As a concrete example, consider the TOP for the mission rehearsal domain. We first specify the team organization hierarchy (see Figure 4(a)). *Task Force* is the highest level team in this organization and consists of two roles *Scouting* and *Transport*, where the *Scouting* sub-team has roles for each of the three scouting sub-sub-teams. Next we specify a hierarchy of reactive team plans (Figure 4(b)). Reactive team plans explicitly express joint activities of the relevant team and consist of: (i) pre-conditions under which the plan is to be proposed; (ii) termination conditions under which the plan is to be ended; and (iii) team-level actions to be executed as part of the plan (an example plan will be discussed shortly). In Figure 4(b), the highest level plan **Execute Mission** has three sub-plans: **DoScouting** to make one path from X to Y safe for the transports, **DoTransport** to move the transports along a scouted path, and **RemainingScouts** for the scouts which have not reached the destination yet to get there.

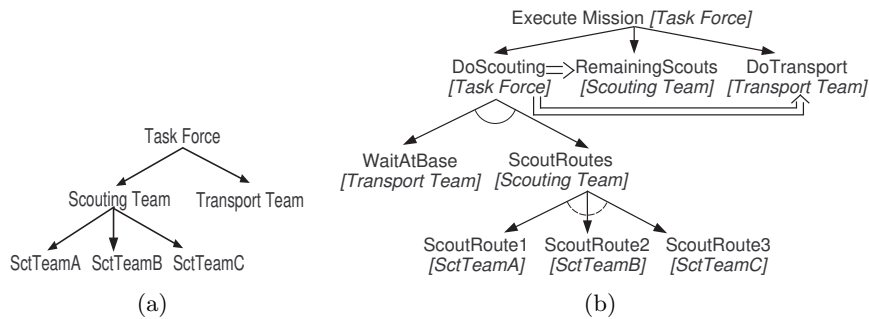

(a)                                    (b)

Figure 4: TOP for mission rehearsal domain a: Organization hierarchy; b: Plan hierarchy.

Figure 4(b) also shows coordination relationships: An AND relationship is indicated with a solid arc, while an OR relationship is indicated with a dashed arc. Thus, **WaitAtBase** and **ScoutRoutes** must both be done while at least one of **ScoutRoute1**, **ScoutRoute2** or **ScoutRoute3** need be performed. There is also a temporal dependence relationship among the sub-plans, which implies that sub-teams assigned to perform **DoTransport** or **RemainingScouts** cannot do so until the **DoScouting** plan has completed. However, **DoTransport** and **RemainingScouts** execute in parallel. Finally, we assign roles to plans – Figure 4(b) shows the assignment in brackets adjacent to the plans. For instance, *Task Force* team is assigned to jointly perform **Execute Mission** while *Sct-TeamA* is assigned to **ScoutRoute1**.

The team plan corresponding to **Execute Mission** is shown in Figure 5. As can be seen, each team plan consists of a context, pre-conditions, post-conditions, body and constraints. The context describes the conditions that must be fulfilled in the parent plan while the pre-conditions are the particular conditions that will cause this sub-plan to begin exe-cution. Thus, for **Execute Mission**, the pre-condition is that the team mutually believes (MB)[1] that they are the "start" location. The post-conditions are divided into Achieved, Unachievable and Irrelevant conditions under which this sub-plan will be terminated. The body consists of sub-plans that exist within this team plan. Lastly, constraints describe any temporal constraints that exist between sub-plans in the body. The description of all the plans in the plan hierarchy of Figure 4(b) is given in Appendix A.

<u>ExecuteMission</u>:
*Context:* ∅
*Pre-conditions:* (MB <TaskForce> location(TaskForce) = START)
*Achieved:* (MB <TaskForce> (Achieved(DoScouting) ∧ Achieved(DoTransport))) ∧ (time
> T ∨ (MB <TaskForce>
         Achieved(RemainingScouts) ∨ (∄ helo ∈ ScoutingTeam, alive(helo) ∧
location(helo) ≠ END)))
*Unachievable:* (MB <TaskForce> Unachievable(DoScouting)) ∨ (MB <TaskForce>
Unachievable(DoTransport) ∧
           (Achieved(RemainingScouts) ∨(∄ helo ∈ ScoutingTeam, alive(helo) ∧
location(helo) ≠ END)))
*Irrelevant:* ∅
*Body:*
DoScouting
DoTransport
RemainingScouts
*Constraints:* DoScouting → DoTransport, DoScouting → RemainingScouts

Figure 5: Example team plan. MB refers to mutual belief.

Just as in HTN (Dix, Muoz-Avila, Nau, & Zhang, 2003; Erol, Hendler, & Nau, 1994), the plan hierarchy of a TOP gives a decomposition of the task into smaller tasks. However, the language of TOPs is richer than the language of early HTN planning (Erol et al., 1994) which contained just simple ordering constraints. As seen in the above example, the plan hierarchy in TOPs can also contain relationships like AND and OR. In addition, just like more recent work in HTN planning (Dix et al., 2003), sub-plans in TOPs can contain pre-conditions and post-conditions, thus allowing for conditional plan execution. The main differences between TOPs and HTN planning are: (i) TOPs contain an organization hierarchy in addition to a plan hierarchy, (ii) the TOP interpreter ensures that the team executes its plans coherently. As seen later, TOPs will be analyzed with all of this expressiveness including conditional execution; however, since our analysis will focus on a fixed time horizon, any loops in the task description will be unrolled up to the time horizon.

---

1. Mutual Belief (Wooldridge, 2002), shown as (MB ⟨*team*⟩ x) in Figure 5, refers to a private belief held by each agent in the team that they each believe that a fact x is true, and that each of the other agents in the team believe that x is true, and that every agent believes that every other agent believes that x is true and so on. Such infinite levels of nesting are difficult to realize in practice. Thus, as in practical BDI implementations, for the purposes of this article, a mutual belief is approximated to be a private belief held by an agent that all the agents in the team believe that x is true.

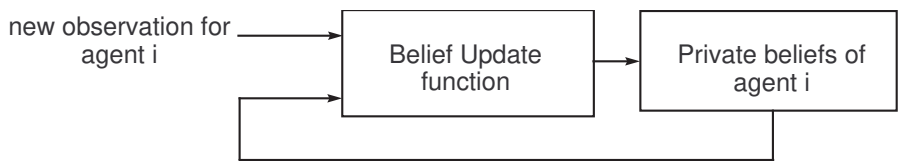

Figure 6: Mapping of observations to beliefs.

During execution, each agent has a copy of the TOP. The agent also maintains a set of private beliefs, which are a set of propositions that the agent believes to be true (see Figure 6). When an agent receives new beliefs, i.e. observations (including communication), the belief update function is used to update its set of privately held beliefs. For instance, upon seeing the last scout crashed, a transport may update its privately held beliefs to include the belief "CriticalFailure(DoScouting)". In practical BDI systems, such belief update computation is of low complexity (e.g. constant or linear time). Once beliefs are updated, an agent selects which plan to execute by matching its beliefs with the pre-conditions in the plans. The basic execution cycle is similar to standard reactive planning systems such as PRS (Georgeff & Lansky, 1986).

During team plan execution, observations in the form of communications often arise because of the coordination actions executed by the TOP interpreter. For instance, TOP interpreters have exploited BDI theories of teamwork, such as Levesque *et al.*'s theory of joint intentions (Levesque, Cohen, & Nunes, 1990) which require that when an agent comes to privately believe a fact that terminates the current team plan (i.e. matches the achievement or unachievability conditions of a team plan), then it communicates this fact to the rest of the team. By performing such coordination actions automatically, the TOP interpreter enables coherence at the initiation and termination of team plans within a TOP. Some further details and examples of TOPs can be seen in the work of Pynadath and Tambe (2003), Tambe *et al.* (2000) and Tidhar (1993b).

We can now more concretely illustrate the key challenges in role allocation mentioned earlier. First, a human developer must allocate available agents to the organization hierarchy (Figure 4(a)), to find the best role allocation. However, there are combinatorially many allocations to choose from (Hunsberger & Grosz, 2000; Tambe et al., 2000). For instance, starting with just 6 homogeneous helicopters results in 84 different ways of deciding how many agents to assign to each scouting and transport sub-team. This problem is exacerbated by the fact that the best allocation varies significantly based on domain variations. For example, Figure 7 shows three different assignments of agents to the team organization hierarchy, each found in our analysis to be the best for a given setting of failure and observation probabilities (details in Section 6). For example, increasing the probability of failures on all routes resulted in the number of transports in the best allocation changing from four (see Figure 7(b)) to three (see Figure 7(a)), where an additional scout was added to *SctTeamB*. If failures were not possible at all, the number of transports increased to five (see Figure 7(c)). Our analysis takes a step towards selecting the best among such allocations.

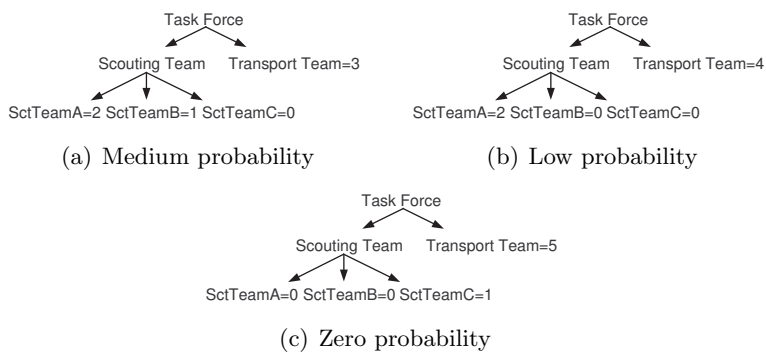

(a) Medium probability          (b) Low probability

(c) Zero probability

Figure 7: Best role allocations for different probabilities of scout failure.

Figure 8 shows the TOP for the RoboCupRescue scenario. As can be seen, the plan hierarchy for this scenario consists of a pair of **ExtinguishFire** and **RescueCivilians** plans done in parallel, each of which further decompose into individual plans. (These individual plans get the fire engines and ambulances to move through the streets using specific search algorithms, however, these individual plans are not relevant for our discussions in this article; interested readers should refer to the description of our RoboCupRescue team entered into the RoboCup competitions of 2001 (Nair, Ito, Tambe, & Marsella, 2002).) The organizational hierarchy consists of *Task Force* comprising of two *Engine* sub-teams, one for each fire and an *Ambulance Team*, where the engine teams are assigned to extinguishing the fires while the ambulance team is assigned to rescuing civilians. In this particular TOP, the assignment of ambulances to *AmbulanceTeamA* and *AmbulanceTeamB* is conditioned on the communication "c", indicated by "AmbulanceTeamA|c" and "AmbulanceTeamB|c". "c" is not described in detail in this figure, but it refers to the communication that is received from the fire engines that describes the number of civilians present at each fire. The problem is which engines to assign to each *Engine Team* and for each possible value of "c", which ambulances to assign to each *Ambulance Team*. Note that engines have differing capabilities owing to differing distances from fires while all the ambulances have identical capabilities.

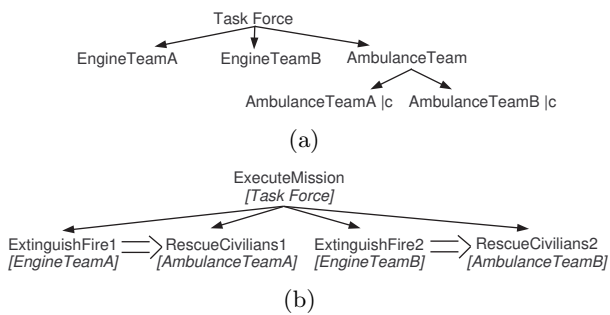

Figure 8: TOP for RoboCupRescue scenario a: Organization hierarchy; b: Plan hierarchy.

## 3. Role-based Multiagent Team Decision Problem

Multiagent Team Decision Problem (MTDP) (Pynadath & Tambe, 2002) is inspired by the economic theory of teams (Marschak & Radner, 1972; Ho, 1980; Yoshikawa, 1978). In order to do quantitative analysis of key coordination decisions in multiagent teams, we extend MTDP for the analysis of the coordination actions of interest. For example, the COM-MTDP (Pynadath & Tambe, 2002) is an extension of MTDP for the analysis of communication. In this article, we illustrate a general methodology for analysis of other aspects of coordination and present the RMTDP model for quantitative analysis of role allocation and reallocation as a concrete example. In contrast to BDI systems introduced in the previous section, RMTDP enables explicit quantitative optimization of team performance. Note that, while we use MTDP, other possible distributed POMDP models could potentially also serve as a basis (Bernstein et al., 2000; Xuan et al., 2001).

### 3.1 Multiagent Team Decision Problem

Given a team of $n$ agents, an MTDP (Pynadath & Tambe, 2002) is defined as a tuple: $\langle S, A, P, \Omega, O, R \rangle$. It consists of a finite set of states $S = \Xi_1 \times \cdots \times \Xi_m$ where each $\Xi_j$, $1 \leq j \leq m$, is a feature of the world state. Each agent $i$ can perform an action from its set of actions $A_i$, where $\times_{1 \leq i \leq n} A_i = A$. $P(s, <a_1, \ldots, a_n>, s')$ gives the probability of transitioning from state $s$ to state $s'$ given that the agents perform the actions $<a_1, \ldots, a_n>$ jointly. Each agent $i$ receives an observation $\omega_i \in \Omega_i$ ($\times_{1 \leq i \leq n} \Omega_i = \Omega$) based on the function $O(s, <a_1, \ldots, a_n>, \omega_1, \ldots, \omega_n)$, which gives the probability that the agents receive the observations, $\omega_1, \ldots, \omega_n$ given that the world state is $s$ and they perform $<a_1, \ldots, a_n>$ jointly. The agents receive a single joint reward $R(s, <a_1, \ldots, a_n>)$ based on the state $s$ and their joint action $<a_1, \ldots, a_n>$. This joint reward is shared equally by all members and there is no other private reward that individual agents receive for their actions. Thus, the agents are motivated to behave as a team, taking the actions that jointly yield the maximum expected reward.

Each agent $i$ in an MTDP chooses its actions based on its local *policy*, $\pi_i$, which is a mapping of its observation history to actions. Thus, at time $t$, agent $i$ will perform action $\pi_i(\omega_i^0, \ldots, \omega_i^t)$. This contrasts with a single-agent POMDP, where we can index an agent's policy by its *belief state* – a probability distribution over the world state (Kaelbling, Littman, & Cassandra, 1998), which is shown to be a *sufficient statistic* in order to compute the optimal policy (Sondik, 1971). Unfortunately, we cannot directly use single-agent POMDP techniques (Kaelbling et al., 1998) for maintaining or updating belief states (Kaelbling et al., 1998) in a MTDP – unlike in a single agent POMDP, in MTDP, an agent's observation depends not only on its own actions, but also on unknown actions of other agents. Thus, as with other distributed POMDP models (Bernstein et al., 2000; Xuan et al., 2001), in MTDP, local policies $\pi_i$ are indexed by observation histories. $\pi = <\pi_1, \ldots, \pi_n>$ refers to the joint policy of the team of agents.

### 3.2 Extension for Explicit Coordination

Beginning with MTDP, the next step in our methodology is to make an explicit separation between domain-level actions and the coordination actions of interest. Earlier work intro-duced the COM-MTDP model (Pynadath & Tambe, 2002), where the coordination action was fixed to be the communication action, and got separated out. However, other coordination actions could also be separated from domain-level actions in order to investigate their impact. Thus, to investigate role allocation and reallocations, actions for allocating agents to roles and to reallocate such roles are separated out. To that end, we define RMTDP (Role-based Multiagent Team Decision Problem) as a tuple $\langle S, A, P, \Omega, O, R, \mathcal{RL} \rangle$ with a new component, $\mathcal{RL}$. In particular, $\mathcal{RL} = \{r_1, \ldots, r_s\}$ is a set of all roles that the agents can undertake. Each instance of role $r_j$ may be assigned some agent $i$ to fulfill it. The actions of each agent are now distinguishable into two types:

**Role-Taking actions:** $\Upsilon_i = \{v_{ir_j}\}$ contains the role-taking actions for agent $i$. $v_{ir_j} \in \Upsilon_i$ means that agent $i$ takes on the role $r_j \in \mathcal{RL}$.

**Role-Execution Actions:** $\Phi_i = \bigcup_{r_j \in \mathcal{RL}} \Phi_{ir_j}$ contains the execution actions for agent $i$ where $\Phi_{ir_j}$ is the set of agent $i$'s actions for executing role $r_j \in \mathcal{RL}$

In addition we define the set of states as $S = \Xi_1 \times \cdots \times \Xi_m \times \Xi_{roles}$, where the feature $\Xi_{roles}$ (a vector) gives the current role that each agent has taken on. The reason for introducing this new feature is to assist us in the mapping from a BDI team plan to an RMTDP. Thus each time an agent performs a new role-taking action successfully, the value of the feature $\Xi_{roles}$ will be updated to reflect this change. The key here is that we not only model an agent's initial role-taking action but also subsequent role reallocation. Modeling both allocation and reallocation is important for an accurate analysis of BDI teams. Note that an agent can observe the part of this feature pertaining to its own current role but it may not observe the parts pertaining to other agents' roles.

The introduction of roles allows us to represent the specialized behaviors associated with each role, e.g. a transport vs. a scout role. While filling a particular role, $r_j$, agent $i$ can perform only role-execution actions, $\phi \in \Phi_{ir_j}$, which may be different from the role-execution actions $\Phi_{ir_l}$ for role $r_l$. Thus, the feature $\Xi_{roles}$ is used to filter actions such that only those role-execution actions that correspond to the agent's current role are permitted. In the worst case, this filtering does not affect the computational complexity (see Theorem 1 below) but in practice, it can significantly improve performance when trying to find the optimal policy for the team, since the number of domain actions that each agent can choose from is restricted by the role that the agent has taken on. Also, these different roles can produce varied effects on the world state (modeled via transition probabilities, $P$) and the team's reward. Thus, the policies must ensure that agents for each role have the capabilities that benefit the team the most.

Just as in MTDP, each agent chooses which action to perform by indexing its local policy $\pi_i$ by its observation history. In the same epoch some agents could be doing role-taking actions while others are doing role-execution actions. Thus, each agent's local policy $\pi_i$ can be divided into local role-taking and role-execution policies such that for all observation histories, $\omega_i^0, \ldots, \omega_i^t$, either $\pi_{i\Upsilon}(\omega_i^0, \ldots, \omega_i^t) = \mathbf{null}$ or $\pi_{i\Phi}(\omega_i^0, \ldots, \omega_i^t) = \mathbf{null}$. $\pi_\Upsilon = \langle \pi_{1\Upsilon}, \ldots, \pi_{n\Upsilon} \rangle$ refers to the joint role-taking policy of the team of agents while $\pi_\Phi = \langle \pi_{1\Phi}, \ldots, \pi_{n\Phi} \rangle$ refers to the joint role-execution policy.

In this article we do not explicitly model communicative actions as a special action. Thus communication is treated like any other role-execution action and the communication received from other agents are treated as observations.[2]

## 3.3 Complexity Results with RMTDP

While Section 2.2 qualitatively emphasized the difficulty of role allocation, RMTDP helps us in understanding the complexity more precisely. The goal in RMTDP is to come up with joint policies $\pi_\Upsilon$ and $\pi_\Phi$ that will maximize the total expected reward over a finite horizon $T$. Note that agents can change their roles according to their local role-taking policies. The agent's role-execution policy subsequent to this change would contain actions pertaining to this new role. The following theorem illustrates the complexity of finding such optimal joint policies.

**Theorem 1** *The decision problem of determining if there exist policies, $\pi_\Upsilon$ and $\pi_\Phi$, for an RMTDP, that yield an expected reward of at least $K$ over some finite horizon $T$ is NEXP-complete.*

**Proof sketch:** Proof follows from the reduction of MTDP (Pynadath & Tambe, 2002) to/from RMTDP. To reduce MTDP to RMTDP, we set RMTDP's role taking actions, $\Upsilon'$, to null and set the RMTDP's role-execution actions, $\Phi'$, to the MTDP's set of actions, $A$. To reduce RMTDP to MTDP, we generate a new MTDP such that its set of actions, $A'$ is equal to $\Upsilon \bigcup \Phi$. Finding the required policy in MTDP is NEXP-complete (Pynadath & Tambe, 2002).□

As this theorem shows us, solving the RMTDP for the optimal joint role-taking and role-execution policies over even a finite horizon is highly intractable. Hence, we focus on the complexity of just determining the optimal role-taking policy, given a *fixed role-execution policy*. By *fixed role-execution policy*, we mean that the action selection of an agent is predetermined by the role it is executing.

**Theorem 2** *The decision problem of determining if there exists a role-taking policy, $\pi_\Upsilon$, for an RMTDP, that yields an expected reward of at least $K$ together with a fixed role-execution policy $\pi_\Phi$, over some finite horizon $T$ is NEXP-complete.*

**Proof sketch:** We reduce an MTDP to an RMTDP with a different role-taking and a role-execution action corresponding to each action in the MTDP. Hence, in the RMTDP we have a role-taking action $v_{ir_j}$ for agent $i$ to take on role $r_j$ created for each action $a_j \in A_i$ in the MTDP and each such role $r_j$ contains a single role-execution action, i.e. $|\Phi_{ir_j}| = 1$. For the RMTDP, construct the transition function to be such that a role-taking action always succeeds and the only affected state feature is $\Xi_{roles}$. For the role-execution action $\phi \in \Phi_{ir_j}$, the transition probability is the same as that of the MTDP action, $a_j \in A_i$ corresponding to the last role-taking action $v_{ir_j}$. The fixed role-execution policy is to simply perform the action, $\phi \in \Phi_{ir_j}$, corresponding to the last successful role-taking action, $v_{ir_j}$. Thus, the decision problem for an RMTDP with a fixed role-execution policy is at least as hard

---

2. For a more explicit analysis of communication please refer to work done by Pynadath and Tambe (2002) and Goldman *et al.* (2003).

as the decision problem for an MTDP. Furthermore, given Theorem 1, we can conclude NEXP-Completeness.□

This result suggests that even by fixing the role-execution policy, solving the RMTDP for the optimal role-taking policy is still intractable. Note that Theorem 2 refers to a completely general *globally optimal* role-taking policy, where any number of agents can change roles at any point in time. Given the above result, in general the globally optimal role-taking policy will likely be of doubly exponential complexity, and so we may be left no choice but to run a brute-force policy search, i.e. to enumerate all the role-taking policies and then evaluate them, which together determine the run-time of finding the globally optimal policy. The number of policies is $\left(|\Upsilon|^{\frac{|\Omega|^T - 1}{|\Omega| - 1}}\right)^n$, i.e. doubly exponential in the number of observation histories and the number of agents. Thus, while RMTDP enables quantitative evaluation of team's policies, computing optimal policies is intractable; furthermore, given its low level of abstraction, in contrast to TOP, it is difficult for a human to understand the optimal policy. This contrast between RMTDP and TOP is at the root of our hybrid model described in the following section.

## 4. Hybrid BDI-POMDP Approach

Having explained TOP and RMTDP, we can now present a more detailed view of our hybrid methodology to quantitatively evaluate a TOP. We first provide a more detailed interpretation of Figure 1. BDI team plans are essentially TOP plans, while the BDI interpreter is the TOP coordination layer. As shown in Figure 1, an RMTDP model is constructed corresponding to the domain and the TOP and its interpreter are converted into a corresponding (incomplete) RMTDP policy. We can then analyze the TOP using analysis techniques that rely on evaluating the RMTDP policy using the RMTDP model of the domain.

Thus, our hybrid approach combines the strengths of the TOPs (enabling humans to specify TOPs to coordinate large-scale teams) with the strengths of RMTDP (enabling quantitative evaluation of different role allocations). On the one hand, this synergistic interaction enables RMTDPs to improve the performance of TOP-based BDI teams. On the other hand, we have identified at least six specific ways in which TOPs make it easier to build RMTDPs and to efficiently search RMTDP policies: two of which are discussed in this section, and four in the next section. In particular, the six ways are:

1. TOPs are exploited in constructing RMTDP models of the domain (Section 4.1);

2. TOPs are exploited to present incomplete policies to RMTDPs, restricting the RMTDP policy search (Section 5.1);

3. TOP belief representation is exploited in enabling faster RMTDP policy evaluation (Section 4.2);

4. TOP organization hierarchy is exploited in hierarchically grouping RMTDP policies (Section 5.1);

5. TOP plan hierarchy is exploited in decomposing RMTDPs (Section 5.3);

6. TOP plan hierarchies are also exploited in cutting down the observation or belief histories in RMTDPs (Section 5.3).

The end result of this efficient policy search is a completed RMTDP policy that improves TOP performance. While we exploit the TOP framework, other frameworks for tasking teams, e.g. Decker and Lesser (1993) and Stone and Veloso (1999) could benefit from a similar synergistic interaction.

## 4.1 Guidelines for Constructing an RMTDP

As shown in Figure 1, our analysis approach uses as input an RMTDP model of the domain, as well as an incomplete RMTDP policy. Fortunately, not only does the TOP serve as a direct mapping to the RMTDP policy, but it can also be utilized in actually constructing the RMTDP model of the domain. In particular, the TOP can be used to determine which domain features are important to model. In addition, the structure in the TOP can be exploited in decomposing the construction of the RMTDP.

The elements of the RMTDP tuple, $\langle S, A, P, \Omega, O, R, \mathcal{RL} \rangle$, can be defined using a procedure that relies on both the TOP as well as the underlying domain. While this procedure is not automated, our key contribution is recognizing the exploitation of TOP structures in constructing the RMTDP model. First, in order to determine the set of states, $S$, it is critical to model the variables tested in the pre-conditions, termination conditions and context of all the components (i.e. sub-plans) in the TOP. Note that a state only needs to model the features tested in the TOP; if a TOP pre-condition expresses a complex test on the feature, that test is not modeled in the state, but instead gets used in defining the incomplete policy input to RMTDP. Next we define the set of roles, $\mathcal{RL}$, as the leaf-level roles in the organization hierarchy of the TOP. Furthermore, as specified in Section 3.2, we define a state feature $\Xi_{roles}$ as a vector containing the current role for each agent. Having defined $\mathcal{RL}$ and $\Xi_{roles}$, we now define the actions, $A$ as follows. For each role $r_j \in \mathcal{RL}$, we define a corresponding role-taking action, $v_{ir_j}$ which will succeed or fail depending on the agent $i$ that performs the action and the state $s$ that the action was performed in. The role-execution actions, $\Phi_{ir_j}$ for agent $i$ in role $r_j$, are those allowed for that role according to the TOP.

Thus, we have defined $S$, $A$ and $\mathcal{RL}$ based on the TOP. To illustrate these steps, consider the plans in Figure 4(b). The pre-conditions of the leaf-level plan **ScoutRoute1** (See Appendix A), for instance, tests start location of the helicopters to be at start location X, while the termination conditions test that scouts are at end location Y. Thus, the locations of the helicopters are modeled as features in the set of states in the RMTDP. Using the organization hierarchy, we define the set of roles $\mathcal{RL}$ with a role corresponding to each of the four different kinds of leaf-level roles, i.e. $\mathcal{RL} = \{memberSctTeamA, memberSctTeamB,$ $memberSctTeamC, memberTransportTeam\}$. The role-taking and role-execution actions can be defined as follows:

- A role-taking action is defined corresponding to each of the four roles in $\mathcal{RL}$, i.e. becoming a member of one of the three scouting teams or of the transport team. The domain specifies that only a transport can change to a scout and thus the role-taking action, $jointTransportTeam$, will fail for agent $i$, if the current role of agent $i$ is a scout.

- Role-execution actions are obtained from the TOP plans corresponding to the agent's role. In the mission rehearsal scenario, an agent, fulfilling a scout role (members of SctTeamA, SctTeamB or SctTeamC), always goes forward, making the current position safe, until it reaches the destination and so the only execution action we will consider is "move-making-safe". An agent in a transport role (members of Transport Team) waits at X until it obtains observation of a signal that one scouting sub-team has reached Y and hence the role-execution actions are "wait" and "move-forward".

We must now define $\Omega, P, O, R$. We obtain the set of observations $\Omega_i$ for each agent $i$ directly from the domain. For instance, the transport helos may observe the status of scout helos (normal or destroyed), as well as a signal that a path is safe. Finally, determining the functions, $P, O, R$ requires some combination of human domain expertise and empirical data on the domain behavior. However, as shown later in Section 6, even an approximate model of transitional and observational uncertainty is sufficient to deliver significant benefits. Defining the reward and transition function may sometimes require additional state variables to be modeled, if they were only implicitly modeled in the TOP. In the mission rehearsal domain, the time at which the scouting and transport mission were completed determined the amount of reward. Thus, time was only implicitly modeled in the TOP and needed to be explicitly modeled in the RMTDP.

Since we are interested in analyzing a particular TOP with respect to uncertainty, the procedure for constructing an RMTDP model can be simplified by exploiting the hierarchical decomposition of the TOP in order to decompose the construction of the RMTDP model. The high-level components of a TOP often represent plans executed by different sub-teams, which may only loosely interact with each other. Within a component, the sub-team members may exhibit a tight interaction, but our focus is on the "loose coupling" across components, where only the end results of one component feed into another, or the components independently contribute to the team goal. Thus, our procedure for constructing an RMTDP exploits this loose coupling between components of the plan hierarchy in order to build an RMTDP model represented as a combination of smaller RMTDPs (factors). Note that if such decomposition is infeasible, our approach still applies except that the benefits of the hierarchical decomposition will be unavailable.

We classify sibling components as being either parallel or sequentially executed (contains a temporal constraint). Components executed in parallel could be either *independent* or *dependent*. For *independent* components, we can define RMTDPs for each of these components such that the sub-team executing one component cannot affect the transitions, observations and reward obtained by the sub-teams executing the other components. The procedure for determining the elements of the RMTDP tuple for component $k$, $\langle S_k, A_k, P_k, \Omega_k, O_k, R_k, \mathcal{RL}_k \rangle$, is identical to the procedure described earlier for constructing the overall RMTDP. However, each such component has a smaller set of relevant variables and roles and hence specifying the elements of its corresponding RMTDP is easier.

We can now combine the RMTDPs of the independent components to obtain the RMTDP corresponding to the higher-level component. For a higher level component $l$, whose child components are independent, the set of states, $S_l = \times_{\forall \Xi_x \in F_{S_l}} \Xi_x$ such that $F_{S_l} = \bigcup_{\forall k \ s.t. \ Child(k,l)=\textbf{true}} F_{S_k}$ where $F_{S_l}$ and $F_{S_k}$ are the sets of features for the set of states $S_l$ and set of states $S_k$. A state $s_l \in S_l$ is said to correspond to the state $s_k \in S_k$ if $\forall \Xi_x \in F_{S_k}, s_l[\Xi_x] = s_k[\Xi_x]$, i.e. the state $s_l$ has the same value as state $s_k$ for all features of state $s_k$. The transition function is defined as follows, $P_l(s'_l, a_l, s_l) = \prod_{\forall k \ s.t. \ Child(k,l)=\text{true}} P_k(s'_k, a_k, s_k)$, where $s_l$ and $s'_l$ of component $l$ corresponds to states $s_k$ and $s'_k$ of component $k$ and $a_k$ is the joint action performed by the sub-team assigned to component $k$ corresponding to the joint action $a_l$ performed by the sub-team assigned to component $l$. The observation function is defined similarly as $O_l(s_l, a_l, \omega_l) = \prod_{\forall k \ s.t. \ Child(k,l)=\text{true}} O_k(s_k, a_k, \omega_k)$. The reward function for component $l$ is defined as $R_l(s_l, a_l) = \sum_{\forall k \ s.t. \ Child(k,l)=\text{true}} R_k(s_k, a_k)$.

In the case of sequentially executed components (those connected by a temporal constraint), the components are loosely coupled since the end states of the preceding component specify the start states of the succeeding component. Thus, since only one component is active at a time, the transition function is defined as follows, $P_l(s'_l, a_l, s_l) = P_k(s'_k, a_k, s_k)$, where component $k$ is the only active child component, $s_k$ and $s'_k$ represent the states of component $k$ corresponding to states $s_l$ and $s'_l$ of component $l$ and $a_k$ is the joint action performed by the sub-team assigned to component $k$ corresponding to the joint action $a_l$ performed by the sub-team corresponding to component $l$. Similarly, we can define $O_l(s_l, a_l, \omega_l) = O_k(s_k, a_k, \omega_k)$ and $R_l(s_l, a_l) = R_k(s_k, a_k)$, where $k$ is the only active child component.

Consider the following example from the mission rehearsal domain where components exhibit both sequential dependence and parallel independence. Concretely, the component **DoScouting** is executed first followed by **DoTransport** and **RemainingScouts**, which are parallel and independent and hence, either **DoScouting** is active or **DoTransport** and **RemainingScouts** are active at any point in the execution. Hence, the transition, observation and reward functions of their parent **Execute Mission** is given by the corresponding functions of either **DoScouting** or by the combination of the corresponding functions of **DoTransport** and **RemainingScouts**.

We use a top-down approach in order to determine how to construct a factored RMTDP from the plan hierarchy. As shown in Algorithm 1, we replace a particular sub-plan by its constituent sub-plans if they are either independent or sequentially executed. If not, then the RMTDP is defined using that particular sub-plan. This process is applied recursively starting at the root component of the plan hierarchy. As a concrete example, consider again our mission rehearsal simulation domain and the hierarchy illustrated in Figure 4(b). Given the temporal constraints between **DoScouting** and **DoTransport**, and **DoScouting** and **RemainingScouts**, we exploited sequential decomposition, while **DoTransport** and **RemainingScouts** were parallel and independent components. Hence, we can replace **ExecuteMission** by **DoScouting**, **DoTransport** and **RemainingScouts**. We then apply the same process to **DoScouting**. The constituent components of **DoScouting** are neither independent nor sequentially executed and thus **DoScouting** cannot be replaced by its constituent components. Thus, RMTDP for the mission rehearsal domain is comprised of smaller RMTDPs for **DoScouting**, **DoTransport** and **RemainingScouts**.

Thus, using the TOP to identify relevant variables and building a factored RMTDP utilizing the structure of TOP to decompose the construction procedure, reduce the load on the domain expert for model construction. Furthermore, as shown in Section 5.3, this factored model greatly improves the performance of the search for the best role allocation.

---

**Algorithm 1** Build-RMTDP(TOP top, Sub-plan subplan)

1: children ← subplan→children() {subplan→children() returns the sub-plans within sub-plan}
2: **if** children = **null** or children are not (loosely coupled or independent) **then**
3:    rmtdp ← Define-RMTDP(subplan) {not automated}
4:    return rmtdp
5: **else**
6:    **for all** child in children **do**
7:       factors[child] ← Build-RMTDP(top,child)
8:    rmtdp ← ConstructFromFactors(factors)
9:    return rmtdp

---

## 4.2 Exploiting TOP Beliefs in Evaluation of RMTDP Policies

We now present a technique for exploiting TOPs in speeding up evaluation of RMTDP policies. Before we explain our improvement, we first describe the original algorithm for determining the expected reward of a joint policy, where the local policies of each agent are indexed by its entire observation histories (Pynadath & Tambe, 2002; Nair, Pynadath, Yokoo, Tambe, & Marsella, 2003a). Here, we obtain an RMTDP policy from a TOP as follows. We obtain $\pi_i(\vec{\omega}_i^t)$, i.e. the action performed by agent $i$ for each observation history $\vec{\omega}_i^t$, as the action $a$ performed by the agent $i$ following the TOP when it has a set of privately held beliefs corresponding to the observation history, $\vec{\omega}_i^t$. We compute the expected reward for the RMTDP by projecting the team's execution over all possible branches on different world states and different observations. At each time step, we can compute the expected value of a joint policy, $\pi = <\pi_1, \ldots, \pi_n>$, for a team starting in a given state, $s^t$, with a given set of past observations, $\vec{\omega}_1^t, \ldots, \vec{\omega}_n^t$, as follows:

$$V_\pi^t(s^t, \langle \vec{\omega}_1^t, \ldots, \vec{\omega}_n^t \rangle) = R(s^t, \langle \pi_1(\vec{\omega}_1^t), \ldots, \pi_n(\vec{\omega}_n^t) \rangle) + \sum_{s^{t+1} \in S} P\left(s^t, \langle \pi_1\left(\vec{\omega}_1^t\right), \ldots, \pi_n\left(\vec{\omega}_n^t\right) \rangle, s^{t+1}\right)$$

$$\sum_{\omega_{t+1} \in \Omega} O\left(s^{t+1}, \langle \pi_1(\vec{\omega}_1^t), \ldots, \pi_n(\vec{\omega}_n^t) \rangle, \langle \omega_1^{t+1}, \ldots, \omega_n^{t+1} \rangle\right) \cdot V_\pi^{t+1}\left(s^{t+1}, \langle \vec{\omega}_1^{t+1}, \ldots, \vec{\omega}_n^{t+1} \rangle\right) \quad (1)$$

The expected reward of a joint policy $\pi$ is given by $V_\pi^0(s^0, < \textbf{null}, \ldots, \textbf{null} >)$ where $s^0$ is the start state. At each time step $t$, the computation of $V_\pi^t$ performs a summation over all possible world states and agent observations and so has a time complexity of $O\left(|S| \cdot |\Omega|\right)$. This computation is repeated for all states and all observation histories of length $t$, i.e. $O\left(|S| \cdot |\Omega|^t\right)$ times. Therefore, given a time horizon $T$, the overall complexity of this algorithm is $O\left(|S|^2 \cdot |\Omega|^{T+1}\right)$.

As discussed in Section 2.2, in a team-oriented program, each agent's action selection is based on just its currently held private beliefs (note that mutual beliefs are modeled as privately held beliefs about all agents as per footnote 2). A similar technique can be exploited when mapping TOP to an RMTDP policy. Indeed, the evaluation of a RMTDP policy that corresponds to a TOP can be speeded up if each agent's local policy is indexed by its private beliefs, $\psi_i^t$. We refer to $\psi_i^t$, as the TOP-congruent belief state of agent $i$ in the RMTDP. Note that this belief state is not a probability distribution over the world states as in a single agent POMDP, but rather the privately held beliefs (from the BDI program) of agent $i$ at time $t$. This is similar to the idea of representing a policy by a finite-state controller (Hansen & Zhou, 2003; Poupart & Boutilier, 2003). In this case, the private beliefs would map to the states of the finite-state controller.

Belief-based RMTDP policy evaluation leads to speedup because multiple observation histories map to the same belief state, $\psi_i{}^t$. This speedup is a key illustration of exploitation of synergistic interactions of TOP and RMTDP. In this instance, belief representation techniques used in TOP are reflected in RMTDP, and the resulting faster policy evaluation can help us optimize TOP performance. A detailed example of belief state is presented later after a brief explanation of how such belief-based RMTDP policies can be evaluated.

Just as with evaluation using observation histories, we compute the expected reward of a belief-based policy by projecting the team's execution over all possible branches on different world states and different observations. At each time step, we can compute the expected value of a joint policy, $\pi = <\pi_1, \ldots, \pi_n>$, for a team starting in a given state, $s^t$, with a given team belief state, $< \psi_1{}^t, \ldots, \psi_n{}^t >$ as follows:

$$V_\pi^t(s^t, \langle \psi_1{}^t \ldots \psi_n{}^t \rangle) = R(s^t, \langle \pi_1(\psi_1{}^t), \ldots, \pi_n(\psi_n{}^t) \rangle) + \sum_{s^{t+1} \in S} P\left(s^t, \langle \pi_1\left(\psi_1{}^t\right), \ldots, \pi_n\left(\psi_n{}^t\right) \rangle, s^{t+1}\right)$$

$$\cdot \sum_{\omega_{t+1} \in \Omega} O\left(s^{t+1}, \langle \pi_1(\psi_1{}^t), \ldots, \pi_n(\psi_n{}^t) \rangle, \langle \omega_1^{t+1}, \ldots, \omega_n^{t+1} \rangle \right) \cdot V_\pi^{t+1}\left(s^{t+1}, \langle \psi_1{}^{t+1}, \ldots, \psi_n{}^{t+1} \rangle\right)$$

$$(2)$$

$$where \ \psi_i{}^{t+1} = \textbf{BeliefUpdateFunction}\left(\psi_i{}^t, \omega_i^{t+1}\right)$$

The complexity of computing this function (expression 2) is $O\left(|S| \cdot |\Omega|\right) \cdot BF$, where $BF$ represents the complexity of the belief update function, **BeliefUpdateFunction**. At each time step the computation of the value function is done for every state and for all possible reachable belief states. Let $|\Psi_i| = \textbf{max}_{1 \leq t \leq T}(|\psi_i^t|)$ represent the maximum number of possible belief states that agent $i$ can be in at any point in time, where $|\psi_i^t|$ is the number of belief states that agent $i$ can be in at $t$. Therefore the complexity of this algorithm is given by $O(|S|^2 \cdot |\Omega| \cdot (|\Psi_1| \cdot \ldots \cdot |\Psi_n|) \cdot T) \cdot BF$. Note that, in this algorithm $T$ is not in the exponent unlike in the algorithm in expression 1. Thus, this evaluation method will give large time savings if: (i) the quantity $(|\Psi_1| \cdot \ldots \cdot |\Psi_n|) \cdot T$ is much less than $|\Omega|^T$ and (ii) the belief update cost is low. In practical BDI systems, multiple observation histories map often onto the same belief state, and thus usually, $(|\Psi_1| \cdot \ldots \cdot |\Psi_n|) \cdot T$ is much less than $|\Omega|^T$. Furthermore, since the belief update function mirrors practical BDI systems, its complexity is also a low polynomial or a constant. Indeed, our experimental results show that significant speedups result from switching to our TOP-congruent belief states $\psi_i{}^t$. However, in the absolute worst case, the belief update function may simply append the new observation to the history of past observations (i.e., TOP-congruent beliefs will be equivalent to keeping entire observation histories) and thus belief-based evaluation will have the same complexity as the observation history-based evaluation.

We now turn to an example of belief-based policy evaluation from the mission rehearsal domain. At each time step, the transport helicopters may receive an observation about whether a scout has failed based on some observation function. If we use the observation-history representation of the policy, then each transport agent would maintain a complete history of the observations that it could receive at each time step. For example, in a setting with two scout helicopters, one on route 1 and the other on route 2, a particular transport helicopter may have several different observation histories of length two. At every time step, the transports may receive an observation about each scout being alive or having failed. Thus, at time $t = 2$, a transport helicopter might have one of the following observation histories of length two, $< \{sct1OnRoute1Alive, sct2OnRoute2Alive\}^1, \{sct1OnRoute1Failed, sct2OnRoute2Failed\}^2 >, < \{sct1OnRoute1Alive, sct2OnRoute2Failed\}^1, \{sct1OnRoute1Failed\}^2 >, < \{sct1OnRoute1Failed, sct2OnRoute2Alive\}^1, \{sct2OnRoute2Failed\}^2 >,$ etc. However, the action selection of the transport helicopters depends on only whether a critical failure (i.e. the last remaining scout has crashed) has taken place to change its role. Whether a failure is critical can be determined by passing each observation through a belief-update function. The exact order in which the observations are received or the precise times at which the failure or non-failure observations are received are not relevant to determining if a critical failure has taken place and consequently whether a transport should change its role to a scout. Thus, many observation histories map onto the same belief states. For example, the above three observation histories all map to the same belief $CriticalFailure(DoScouting)$ i.e. a critical failure has taken place. This results in significant speedups using belief-based evaluation, as Equation 2 needs to be executed over a smaller number of belief states, linear in $T$ in our domains, as opposed to the observation history-based evaluation, where Equation 1 is executed over an exponential number of observation histories ($|\Omega|^T$). The actual speedup obtained in the mission rehearsal domain is demonstrated empirically in Section 6.

## 5. Optimizing Role Allocation

While Section 4 focused on mapping a domain of interest onto RMTDP and algorithms for policy evaluation, this section focuses on efficient techniques for RMTDP policy search, in service of improving BDI/TOP team plans. The TOP in essence provides an incomplete, fixed policy, and the policy search optimizes decisions left open in the incomplete policy; the policy thus completed optimizes the original TOP (see Figure 1). By enabling the RMTDP to focus its search on incomplete policies, and by providing ready-made decompositions, TOPs assist RMTDPs in quickly searching through the policy space, as illustrated in this section. We focus, in particular, on the problem of role allocation (Hunsberger & Grosz, 2000; Modi, Shen, Tambe, & Yokoo, 2003; Tidhar et al., 1996; Fatima & Wooldridge, 2001), a critical problem in teams. While the TOP provides an incomplete policy, keeping open the role allocation decision for each agent, the RMTDP policy search provides the optimal role-taking action at each of the role allocation decision points. In contrast to previous role allocation approaches, our approach determines the best role allocation, taking into consideration the uncertainty in the domain and future costs. Although demonstrated for solving the role allocation problem, the methodology is general enough to apply to other coordination decisions.

## 5.1 Hierarchical Grouping of RMTDP Policies

As mentioned earlier, to address role allocation, the TOP provides a policy that is complete, except for the role allocation decisions. RMTDP policy search then optimally fills in the role allocation decisions. To understand the RMTDP policy search, it is useful to gain an understanding of the role allocation search space. First, note that role allocation focuses on deciding how many and what types of agents to allocate to different roles in the organization hierarchy. This role allocation decision may be made at time $t = 0$ or it may be made at a later time conditioned on available observations. Figure 9 shows a partially expanded role allocation space defined by the TOP organization hierarchy in Figure 4(a) for six helicopters. Each node of the role allocation space completely specifies the allocation of agents to roles at the corresponding level of the organization hierarchy (ignore for now, the number to the right of each node). For instance, the root node of the role allocation space specifies that six helicopters are assigned to the *Task Force* (level one) of the organization hierarchy while the leftmost leaf node (at level three) in Figure 9 specifies that one helicopter is assigned to *SctTeamA*, zero to *SctTeamB*, zero to *SctTeamC* and five helicopters to *Transport Team*. Thus, as we can see, each leaf node in the role allocation space is a complete, valid role allocation of agents to roles in the organization hierarchy.

In order to determine if one leaf node (role allocation) is superior to another we evaluate each using the RMTDP by constructing an RMTDP policy for each. In this particular example, the role allocation specified by the leaf node corresponds to the role-taking actions that each agent will execute at time $t = 0$. For example, in the case of the leftmost leaf in Figure 9, at time $t = 0$, one agent (recall from Section 2.2 that this is a homogeneous team and hence which specific agent does not matter) will become a member of *SctTeamA* while all other agents will become members of *Transport Team*. Thus, for one agent $i$, the role-taking policy will include $\pi_{i\Upsilon}(\textbf{null}) = joinSctTeamA$ and for all other agents, $j, j \neq i$, it will include $\pi_{j\Upsilon}(\textbf{null}) = joinTransportTeam$. In this case, we assume that the rest of the role-taking policy, i.e. how roles will be reallocated if a scout fails, is obtained from the role reallocation algorithm in the BDI/TOP interpreter, such as the *STEAM* algorithm (Tambe et al., 2000). Thus for example, if the role reallocation is indeed performed by the STEAM algorithm, then STEAM's reallocation policy is included into the incomplete policy that the RMTDP is initially provided. Thus, the best role allocation is computed keeping in mind STEAM's reallocation policy. In STEAM, given a failure of an agent playing $Role_F$, an agent playing $Role_R$ will replace it if:

$$Criticality(Role_F) - Criticality(Role_R) > 0$$
$$Criticality(x) = 1 \; if \; x \; is \; critical; = 0 \; otherwise$$

Thus, if based on the agents' observations, a critical failure has taken place, then the replacing agent's decision to replace or not will be computed using the above expression and then included in the incomplete policy input to the RMTDP. Since such an incomplete policy is completed by the role allocation at each leaf node using the technique above, we have been able to construct a policy for the RMTDP that corresponds to the role allocation.

In some domains like RoboCupRescue, not all allocation decisions are made at time $t = 0$. In such domains, it is possible for the role allocation to be conditioned on observations (or communication) that are obtained during the course of the execution. For instance, as shown in Figure 8(a), in the RoboCupRescue scenario, the ambulances are allocated to the sub-team *AmbulanceTeamA* or *AmbulanceTeamB* only after information about the location of civilians is conveyed to them by the fire engines. The allocation of the ambulances is then conditioned on this communication, i.e. on the number of civilians at each location. Figure 10 shows the partially expanded role allocation for a scaled-down rescue scenario with three civilians, two ambulances and two fire engines (one at station 1 and the other at station 2). In the Figure, 1;1;2 depicts the fact that there are two ambulances, while there is one fire engine at each station. As shown, there is a level for the allocation of fire engines to *EngineTeamA* and *EngineTeamB* which gives the number of engines assigned to each *EngineTeam* from each station. The next level (leaf level) has different leaf nodes for each possible assignment of ambulances to *AmbulanceTeamA* and *AmbulanceTeamB* depending upon the value of communication "c". Since there are three civilians and we exclude the case where no civilians are present at a particular fire, there are two possible messages i.e. one civilian at fire 1 or two civilians at fire 1 ($c = 1$ or 2).

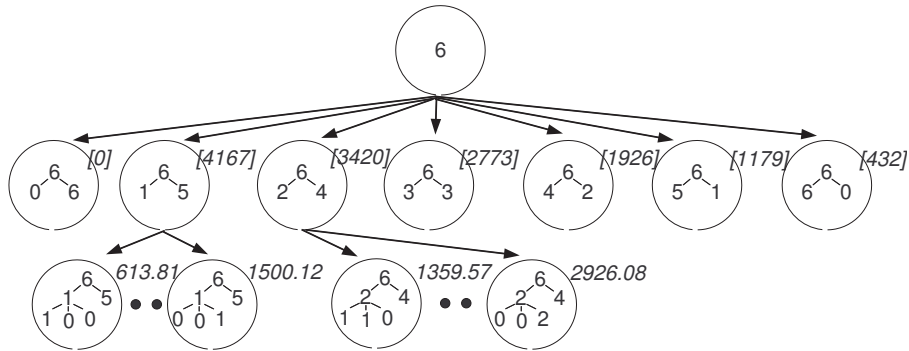

Figure 9: Partially expanded role allocation space for mission rehearsal domain(six helos).

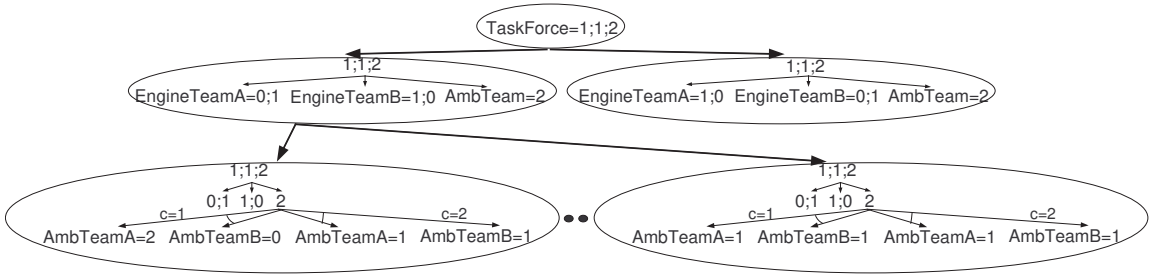

Figure 10: Partially expanded role allocation space for Rescue domain (one fire engine at station 1, one fire engine at station 2, two ambulances, three civilians).

We are thus able to exploit the TOP organization hierarchy to create a hierarchical grouping of RMTDP policies. In particular, while the leaf node represents a complete RMTDP policy (with the role allocation as specified by the leaf node), a parent node represents a group of policies. Evaluating a policy specified by a leaf node is equivalent to evaluating a specific role allocation while taking future uncertainties into account. We could do a brute force search through all role allocations, evaluating each in order to determine the best role allocation. However, the number of possible role allocations is exponential in the leaf roles in the organization hierarchy. Thus, we must prune the search space.

## 5.2 Pruning the Role Allocation Space

We prune the space of valid role allocations using upper bounds (MaxEstimates) for the parents of the leaves of the role allocation space as admissible heuristics (Section 5.3). Each leaf in the role allocation space represents a completely specified policy and the MaxEstimate is an upper bound of maximum value of all the policies under the same parent node evaluated using the RMTDP. Once we obtain MaxEstimates for all the parent nodes (shown in brackets to the right of each parent node in Figure 9), we use branch-and-bound style pruning (see Algorithm 2). While we discuss Algorithm 2 below, we note that in essence it performs branch-and-bound style pruning; the key novelty is step 2 which we discuss in Section 5.3.

The branch-and-bound algorithm works as follows: First, we sort the parent nodes by their estimates and then start evaluating children of the parent with the highest MaxEstimate (Algorithm 2: steps 3-13). Evaluate(RMTDP, child) refers to the evaluation of the leaf-level policy, child, using the RMTDP model. This evaluation of leaf-level policies (step 13) can be done using either of the methods described in Section 4. In the case of the role allocation space in Figure 9, we would start with evaluating the leaves of the parent node that has one helicopter in *Scouting Team* and five in *Transport Team*. The value of evaluating each leaf node is shown to the right of the leaf node. Once we have obtained the value of the best leaf node (Algorithm 2: steps 14,15), in this case 1500.12, we compare this with the MaxEstimates of the other parents of the role allocation space (Algorithm 2: steps 16-18). As we can see from Figure 9 this would result in pruning of three parent nodes (leftmost parent and right two parents) and avoid the evaluation of 65 of the 84 leaf-level policies. Next, we would then proceed to evaluate all the leaf nodes under the parent with two helos in *Scouting Team* and four in *Transport Team*. This would result in pruning of all the remaining unexpanded parent nodes and we will return the leaf with the highest value, which in this case is the node corresponding to two helos allocated to *SctTeamA* and four to *Transport Team*. Although demonstrated for a 3-level hierarchy, the methodology for applying to deeper hierarchies is straightforward.

## 5.3 Exploiting TOP to Calculate Upper Bounds for Parents

We will now discuss how the upper bounds of parents, called MaxEstimates, can be calculated for each parent. The MaxEstimate of a parent is defined as a strict upper bound of the maximum of the expected reward of all the leaf nodes under it. It is necessary that the MaxEstimate be an upper bound or else we might end up pruning potentially useful role allocations. In order to calculate the MaxEstimate of each parent we could evaluate each of the leaf nodes below it using the RMTDP, but this would nullify the benefit of any subsequent pruning. We, therefore, turn to the TOP plan hierarchy (see Figure 4(b)) to break up this evaluation of the parent node into components, which can be evaluated separately thus decomposing the problem. In other words, our approach exploits the structure of the BDI program to construct small-scale RMTDPs unlike other decomposition techniques which

---

**Algorithm 2** Branch-and-bound algorithm for policy search.

---

1: Parents ← list of parent nodes
2: Compute MAXEXP(Parents) {Algorithm 3}
3: Sort Parents in decreasing order of MAXEXP
4: bestVal ← −∞
5: **for all** parent ∈ Parents **do**
6:     done[parent] ← false; pruned[parent] ← false
7: **for all** parent ∈ Parents **do**
8:     **if** done[parent] = false and pruned[parent] = false **then**
9:         child ← parent→nextChild() {child is a leaf-level policy under parent}
10:         **if** child = null **then**
11:             done[parent] ← true
12:         **else**
13:             childVal ← Evaluate(RMTDP,child)
14:             **if** childVal > bestVal **then**
15:                 bestVal ← childVal;best ← child
16:                 **for all** parent1 in Parents **do**
17:                     **if** MAXEXP[parent1] < bestVal **then**
18:                         pruned[parent1] ← true
19: **return** best

---

just assume decomposition or ultimately rely on domain experts to identify interactions in the agents' reward and transition functions (Dean & Lin, 1995; Guestrin, Venkataraman, & Koller, 2002).

For each parent in the role allocation space, we use these small-scale RMTDPs to evaluate the values for each TOP component. Fortunately, as discussed in Section 4.1, we exploited small-scale RMTDPs corresponding to TOP components in constructing larger scale RMTDPs. We put these small-scale RMTDPs to use again, evaluating policies within each component to obtain upper bounds. Note that just like in evaluation of leaf-level policies, the evaluation of components for the parent node can be done using either the observation histories (see Equation 1) or belief states (see Equation 2). We will describe this section using the observation history-based evaluation method for computing the values of the components of each parent, which can be summed up to obtain its MaxEstimate (an upper bound on its children's values). Thus, whereas a parent in the role allocation space represents a group of policies, the TOP components (sub-plans) allow a component-wise evaluation of such a group to obtain an upper bound on the expected reward of any policy within this group.

Algorithm 3 exploits the smaller-scale RMTDP components, discussed in Section 4.1, to obtain upper bounds of parents. First, in order to evaluate the MaxEstimate for each parent node in the role allocation space, we identify the start states for each component from which to evaluate the RMTDPs. We explain this step using a parent node from Figure 9 – *Scouting Team* = two helos, *Transport Team* = four helos (see Figure 11). For the very first component which does not have any preceding components, the start states corresponds to the start states of the policy that the TOP was mapped onto. For each of the next

components – where the next component is one linked by a sequential dependence – the start states are the end states of the preceding component. However, as explained later in this section, we can significantly reduce this list of start states from which each component can be evaluated.

**Algorithm 3** MAXEXP method for calculating upper bounds for parents in the role allocation space.

1: **for all** parent in search space **do**
2:    MAXEXP[parent] ← 0
3:    **for all** component $i$ corresponding to factors in the RMTDP from Section 4.1 **do**
4:       **if** component $i$ has a preceding component $j$ **then**
5:          Obtain start states, $states[i] \leftarrow endStates[j]$
6:          $states[i] \leftarrow$ **removeIrrelevantFeatures**$(states[i])$ {discard features not present in $S_i$}
7:          Obtain corresponding observation histories at start $OHistories[i] \leftarrow endOHistories[j]$
8:          $OHistories[i] \leftarrow$ **removeIrrelevantObservations**$(OHistories[i])$
9:       **else**
10:         Obtain start states, $states[i]$
11:         Observation histories at start $OHistories[i] \leftarrow$ **null**
12:       $maxEval[i] \leftarrow 0$
13:       **for all** leaf-level policies $\pi$ under parent **do**
14:          $maxEval[i] \leftarrow$ **max**$(maxEval[i], $**max**$_{s_i \in states[i], oh_i \in OHistories[i]}(Evaluate(RMTDP_i, s_i, oh_i, \pi)))$
15:    MAXEXP[parent] $\overset{+}{\leftarrow} maxEval[i]$

Similarly, the starting observation histories for a component are the observation histories on completing the preceding component (no observation history for the very first component). BDI plans do not normally refer to entire observation histories but rely only on key beliefs which are typically referred to in the pre-conditions of the component. Each starting observation history can be shortened to include only these relevant observations, thus obtaining a reduced list of starting observation sequences. Divergence of private observations is not problematic, e.g. will not cause agents to trigger different team plans. This is because as indicated earlier in Section 2.2, TOP interpreters guarantee coherence in key aspects of observation histories. For instance, as discussed earlier, TOP interpreter ensures coherence in key beliefs when initiating and terminating team plans in a TOP; thus avoiding such divergence of observation histories.

In order to compute the maximum value for a particular component, we evaluate all possible leaf-level policies within that component over all possible start states and observation histories and obtain the maximum (Algorithm 3:steps 13-14). During this evaluation, we store all the end states and ending observation histories so that they can be used in the evaluation of subsequent components. As shown in Figure 11, for the evaluation of **DoScouting** component for the parent node where there are two helicopters assigned to *Scouting Team* and four helos to *Transport Team*, the leaf-level policies correspond to all possible ways these helicopters could be assigned to the teams *SctTeamA, SctTeamB, Sct-*

*TeamC* and *Transport Team*, e.g. one helo to *SctTeamB*, one helo to *SctTeamC* and four helos to *Transport Team*, or two helos to *SctTeamA* and four helos to *Transport Team*, etc. The role allocation tells the agents what role to take in the first step. The remainder of the role-taking policy is specified by the role replacement policy in the TOP infrastructure and role-execution policy is specified by the **DoScouting** component of the TOP.

To obtain the MaxEstimate for a parent node of the role allocation space, we simply sum up the maximum values obtained for each component (Algorithm 3:steps 15), e.g. the maximum values of each component (see right of each component in Figure 11) were summed to obtain the MaxEstimate (84 + 3330 + 36 = 3420). As seen in Figure 9, third node from the left indeed has an upper bound of 3420.

The calculation of the MaxEstimate for a parent nodes should be much faster than evaluating the leaf nodes below it in most cases for two reasons. Firstly, parent nodes are evaluated component-wise. Thus, if multiple leaf-level policies within one component result in the same end state, we can remove duplicates to get the start states of the next component. Since each component only contains the state features relevant to it, the number of duplicates is greatly increased. Such duplication of the evaluation effort cannot be avoided for leaf nodes, where each policy is evaluated independently from start to finish. For instance, in the **DoScouting** component, the role allocations, *SctTeamA=1, SctTeamB=1, SctTeamC=0, TransportTeam=4* and *SctTeamA=1, SctTeamB=0, SctTeamC=1, TransportTeam=4*, will have end states in common after eliminating irrelevant features when the scout in *SctTeamB* for the former allocation and the scout in *SctTeamC* for the latter allocation fail. This is because through feature elimination (Algorithm 3:steps 6), the only state features retained for **DoTransport** are the scouted route and number of transports (some transports may have replaced failed scouts) as shown in Figure 11.

The second reason computation of MaxEstimates for parents is much faster is that the number of starting observation sequences will be much less than the number of ending observation histories of the preceding components. This is because not all the observations in the observation histories of a component are relevant to its succeeding components (Algorithm 3:steps 8). Thus, the function **removeIrrelevantObservations** reduces the number of starting observation histories from the observation histories of the preceding component.

We refer to this methodology of obtaining the MaxEstimates of each parent as MAX-EXP. A variation of this, the maximum expected reward with no failures (NOFAIL), is obtained in a similar fashion except that we assume that the probability of any agent failing is 0. We are able to make such an assumption in evaluating the parent node, since we focus on obtaining upper bounds of parents, and not on obtaining their exact value. This will result in less branching and hence evaluation of each component will proceed much quicker. The NOFAIL heuristic only works if the evaluation of any policy without failures occurring is higher than the evaluation of the same policy with failures possible. This should normally be the case in most domains. The evaluation of the NOFAIL heuristics for the role allocation space for six helicopters is shown in square brackets in Figure 9.

The following theorem shows that the MAXEXP method for finding the upper bounds indeed finds an upper bound and thus yields an admissible search heuristic for the branch-and-bound search of the role allocation space.

**Theorem 3** *The MAXEXP method will always yield an upper bound.*

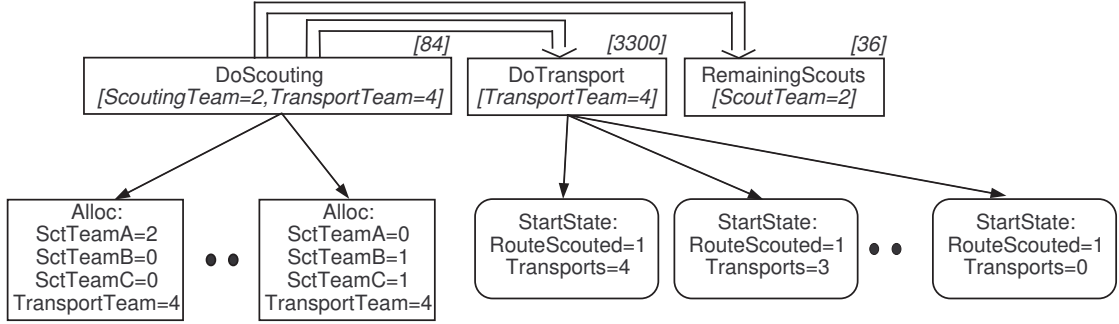

Figure 11: Component-wise decomposition of a parent by exploiting TOP.

**Proof:** See Appendix C.

From Theorem 3, we can conclude that our branch-and-bound policy search algorithm will always find the best role allocation, since the MaxEstimates of the parents are true upper bounds. Also, with the help of Theorem 4, we show that in the worst case, our branch-and-bound policy search has the same complexity as doing a brute force search.

**Theorem 4** *Worst-case complexity for evaluating a single parent node using MAXEXP is the same as that of evaluating every leaf node below it within a constant factor.*

**Proof sketch:**

- The worst case complexity for MAXEXP arises when:

  1. Let $ES_{j\pi}$ be the end states of component $j$ executing policy $\pi$ after removing features that are irrelevant to the succeeding component $k$. Similarly, let $ES_{j\pi'}$ be the end states of component $j$ executing policy $\pi'$ after removing features that are irrelevant to the succeeding component $k$. If $ES_{j\pi} \bigcap ES_{j\pi'} = $ **null** then no duplication in the end states will occur.

  2. Let $OH_{j\pi}$ be the ending observation histories of component $j$ executing policy $\pi$ after removing observations that are irrelevant to the succeeding component $k$. Similarly, let $OH_{j\pi'}$ be the ending observation histories of component $j$ executing policy $\pi'$ after removing observation histories that are irrelevant to the succeeding component $k$. If $OH_{j\pi} \bigcap OH_{j\pi'} = $ **null** then no duplication in the observation histories will occur. Note that if the belief-based evaluation was used then we would replace observation histories by the TOP congruent belief states (see Sect 4).

- In such a case, there is no computational advantage to evaluating each component's MaxEstimate separately. Thus, it is equivalent to evaluating each child node of the parent. Thus, in the worst case, MAXEXP computation for the parent is the same as the evaluating all its children within a constant factor. □

In addition, in the worst case, no pruning will result using MAXEXP and each and every leaf node will need to be evaluated. This is equivalent to evaluating each leaf node twice.

Thus, the worst case complexity of doing the branch-and-bound search using MAXEXP is the same as that of finding the best role allocation by evaluating every leaf node. We refer to this brute-force approach as NOPRUNE. Thus, the worst case complexity of MAXEXP is the same as NOPRUNE. However, owing to pruning and the savings through decomposition in the computation of MaxEstimates, significant savings are likely in the average case. Section 6 highlights these savings for the mission rehearsal and the RoboCupRescue domains.

## 6. Experimental Results

This section presents four sets of results in the context of the two domains introduced in Section 2.1, viz. mission rehearsal and RoboCupRescue (Kitano et al., 1999). First, we investigated empirically the speedups that result from using the TOP-congruent belief states $\psi_i$ (belief-based evaluation) over observation history-based evaluation and from using the algorithm from Section 5 over a brute-force search. Here we focus on determining the best assignment of agents to roles; but assume a fixed TOP and TOP infrastructure. Second, we conducted experiments to investigate the benefits of considering uncertainty in determining role allocations. For this, we compared the allocations found by the RMTDP role allocation algorithm with (i) allocations which do not consider any kind of uncertainty, and (ii) allocations which do not consider observational uncertainty but consider action uncertainty. Third, we conducted experiments in both domains to determine the sensitivity of the results to changes in the model. Fourth, we compare the performance of allocations found by the RMTDP role allocation algorithm with allocations of human subjects in the more complex of our domains – RoboCupRescue simulations.

### 6.1 Results in Mission Rehearsal Domain

For the mission rehearsal domain, the TOP is the one discussed in Section 2.2. As can be seen in Figure 4(a), the organization hierarchy requires determining the number of agents to be allocated to the three scouting sub-teams and the remaining helos must be allocated to the transport sub-team. Different numbers of initial helicopters were attempted, varying from three to ten. The details on how the RMTDP is constructed for this domain are given Appendix B. The probability of failure of a scout at each time step on routes 1, 2 and 3 are 0.1, 0.15 and 0.2, respectively. The probability of a transport observing an alive scout on routes 1, 2 and 3 are 0.95, 0.94 and 0.93, respectively. False positives are not possible, i.e. a transport will not observe a scout as being alive if it has failed. The probability of a transport observing a scout failure on routes 1, 2 and 3 are 0.98, 0.97 and 0.96, respectively. Here too, false positives are not possible and hence a transport will not observe a failure unless it has actually taken place.

Figure 12 shows the results of comparing the different methods for searching the role allocation space. We show four methods. Each method adds new speedup techniques to the previous:

1. NOPRUNE-OBS: A brute force evaluation of every role allocation to determine the best. Here, each agent maintains its complete observation history and the evaluation algorithm in Equation 1 is used. For ten agents, the RMTDP is projected to have in the order of 10,000 reachable states and in the order of 100,000 observation histories per role allocation evaluated (thus the largest experiment in this category was limited to seven agents).

2. NOPRUNE-BEL: A brute force evaluation of every role allocation. The only difference between this method and NOPRUNE-OBS is the use of the belief-based evaluation algorithm (see Equation 2).

3. MAXEXP: The branch-and-bound search algorithm described in Section 5.2 that uses upper bounds of the evaluation of the parent nodes to find the best allocation. Evaluation of the parent and leaf nodes uses the belief-based evaluation.

4. NOFAIL: The modification to branch-and-bound heuristic mentioned in Section 5.3. In essence it is same as MAXEXP, except that the upper bounds are computed making the assumption that agents do not fail. This heuristic is correct in those domains where the total expected reward with failures is always less than if no failures were present and will give significant speedups if agent failures is one of the primary sources of stochasticity. In this method, too, the evaluation of the parent and leaf nodes uses the belief-based evaluation. (Note that only upper bounds are computed using the no-failure assumption – no changes are assumed in the actual domains.)

In Figure 12(a), the Y-axis is the number of nodes in the role allocation space evaluated (includes leaf nodes as well as parent nodes), while in Figure 12(b) the Y-axis represents the runtime in seconds on a *logarithmic* scale. In both figures, we vary the number of agents on the X-axis. Experimental results in previous work using distributed POMDPs are often restricted to just two agents; by exploiting hybrid models, we are able to vary the number of agents from three to ten as shown in Figure 12(a). As clearly seen in Figure 12(a), because of pruning, significant reductions are obtained by MAXEXP and NOFAIL over NOPRUNE-BEL in terms of the numbers of nodes evaluated. This reduction grows quadratically to about 10-fold at ten agents.[3] NOPRUNE-OBS is identical to NOPRUNE-BEL in terms of number of nodes evaluated, since in both methods all the leaf-level policies are evaluated, only the method of evaluation differs. It is important to note that although NOFAIL and MAXEXP result in the same number of nodes being evaluated for this domains, this is not necessarily true always. In general, NOFAIL will evaluate at least as many nodes as MAXEXP since its estimate is at least as high as the MAXEXP estimate. However, the upper bounds are computed quicker for NOFAIL.

Figure 12(b) shows that the NOPRUNE-BEL method provides a significant speedup over NOPRUNE-OBS in actual run-time. For instance, there was a 12-fold speedup using NOPRUNE-BEL instead of NOPRUNE-OBS for the seven agent case (NOPRUNE-OBS could not be executed within a day for problem settings with greater than seven agents). This empirically demonstrates the computational savings possible using belief-based evaluation instead of observation history-based evaluation (see Section 4). For this reason, we use only belief-based evaluation for the MAXEXP and NOFAIL approaches and also for all the remaining experiments in this paper. MAXEXP heuristic results in a 16-fold speedup over NOPRUNE-BEL in the eight agent case.

---

3. The number of nodes for NOPRUNE up to eight agents were obtained from experiments, the rest can be calculated using the formula $[m]^n/n! = (m + n - 1) \cdot \ldots \cdot m/n!$, where $m$ represents the number of heterogeneous role types and $n$ is the number of homogeneous agents. $[m]^n = (m + n - 1) \cdot \ldots \cdot m$ is referred to as a rising factorial.

The NOFAIL heuristic which is very quick to compute the upper bounds far outperforms the MAXEXP heuristic (47-fold speedup over MAXEXP for ten agents). Speedups of MAXEXP and NOFAIL continually increase with increasing number of agents. The speedup of the NOFAIL method over MAXEXP is so marked because, in this domain, ignoring failures results in much less branching.

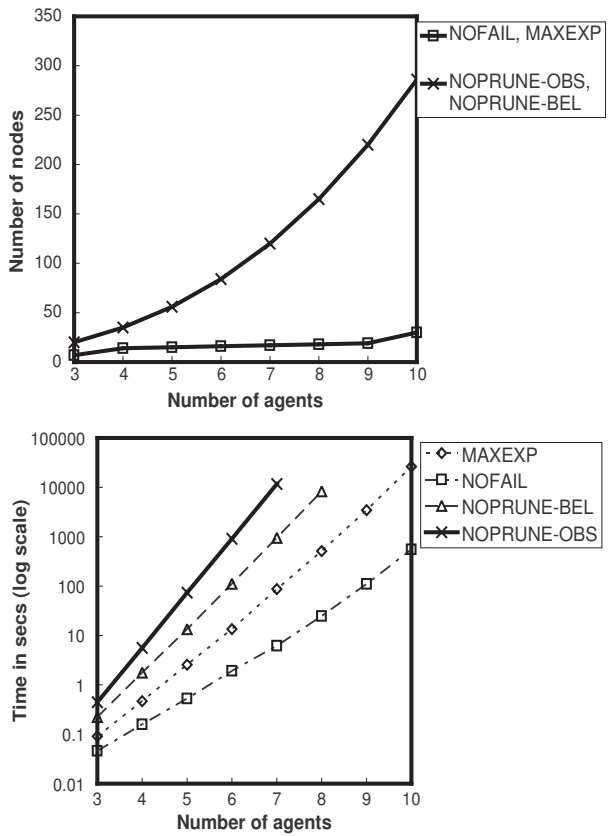

Figure 12: Performance of role allocation space search in mission rehearsal domain, a) (left) Number of nodes evaluated, b) (right)Run-time in seconds on a log scale.

Next, we conducted experiments illustrating the importance of RMTDP's reasoning about action and observation uncertainties on role allocations. For this, we compared the allocations found by the RMTDP role allocation algorithm with allocations found using two different methods (see Figure 13):

1. Role allocation via constraint optimization (COP) (Modi et al., 2003; Mailler & Lesser, 2004) allocation approach: In the COP approach[4], leaf-level sub-teams from the or-ganization hierarchy are treated as variables and the number of helicopters as the domain of each such variable (thus, the domain may be 1, 2, 3,..helicopters). The reward for allocating agents to sub-teams is expressed in terms of constraints:

- Allocating a helicopter to scout a route was assigned a reward corresponding to the route's distance but ignoring the possibility of failure (i.e. ignoring transition probability). Allocating more helicopters to this subteam obtained proportionally higher reward.

- Allocating a helicopter a transport role was assigned a large reward for transporting cargo to the destination. Allocating more helicopters to this subteam obtained proportionally higher reward.

- Not allocating at least one scout role was assigned a reward of negative infinity

- Exceeding the total number of agents was assigned a reward of negative infinity

2. RMTDP with complete observability: In this approach, we consider the transition probability, but ignore partial observability; achieved by assuming complete observability in the RMTDP. An MTDP with complete observability is equivalent to a Markov Decision Problem (MDP) (Pynadath & Tambe, 2002) where the actions are joint actions. We, thus, refer to this allocation method as the MDP method.

---

4. Modi *et al.*'s work (2003) focused on decentralized COP, but in this investigation our emphasis is on the resulting role allocation generated by the COP, and not on the decentralization per se.

Figure 13(a) shows a comparison of the RMTDP-based allocation with the MDP allocation and the COP allocation for increasing number of helicopters (X-axis). We compare using the expected number of transports that get to the destination (Y-axis) as the metric for comparison since this was the primary objective of this domain. As can be seen, considering both forms of uncertainty (RMTDP) performs better than just considering transition uncertainty (MDP) which in turn performs better than not considering uncertainty (COP). Figure 13(b) shows the actual allocations found by the three methods with four helicopters and with six helicopters. In the case of four helicopters (first three bars), RMTDP and MDP are identical, two helicopters scouting route 2 and two helicopters taking on transport role. The COP allocation however consists of one scout on route 3 and three transports. This allocation proves to be too myopic and results in fewer transports getting to the destination safely. In the case of six helicopters, COP chooses just one scout helicopter on route 3, the shortest route. The MDP approach results in two scouts both on route 1, which was longest route albeit the safest. The RMTDP approach, which also considers observational uncertainty chooses an additional scout on route 2, in order to take care of the cases where failures of scouts go undetected by the transports.

It should be noted that the performance of the RMTDP-based allocation will depend on the values of the elements of the RMTDP model. However, as our next experiment revealed, getting the values exactly correct is not necessary. In order to test the sensitivity of the performance of the allocations to the actual model values, we introduced error in the various parameters of the model to see how the allocations found using the incorrect model would perform in the original model (without any errors). This emulates the situation where the model does not correctly represent the domain. Figure 14 shows the expected number of transports that reach the destination (Y-axis) in the mission rehearsal scenario with six helicopters as error (X-axis) is introduced to various parameters in the model. For instance, when the percentage error in failure rate on route 1 (route1-failure-rate) was between -15% (i.e. erroneous failure rate is 85% of actual failure rate) and 10%, there was no difference in the number of transports that reached their destination (3.498). However when the percentage error was greater than 10%, the allocation found was too conservative resulting in fewer transports getting to the destination. Similarly, when the percentage error was less than -15%, the allocation found was too risky, with too few scouts assigned, resulting in more failures. In general, Figure 14 shows that the model is insensitive to errors of 5 to 10% in the model parameters for the mission rehearsal domain, but if the model parameters were outside this range, non-optimal allocations would result. In comparing these non-optimal allocations with COP, we find that they always perform better than COP for the range of errors tested (+/-25%) for both failure rate as well as observability of routes. For instance, at an error of 25% in failure rate on route 1, RMTDP managed to have 2.554 transports safely reach the destination, and COP only managed to get 1.997 transports reach safely. In comparing the non-optimal allocations with MDP, we also find that they performed better than MDP within the range of +/- 25% for error in the observability of the routes. Thus, although the allocations found using an incorrect model were non-optimal they performed better than COP and MDP for large ranges of errors in the model. This shows that getting the model exactly correct is not necessary to find good allocations. We are thus able to obtain benefits from RMTDP even without insisting on an accurate model.

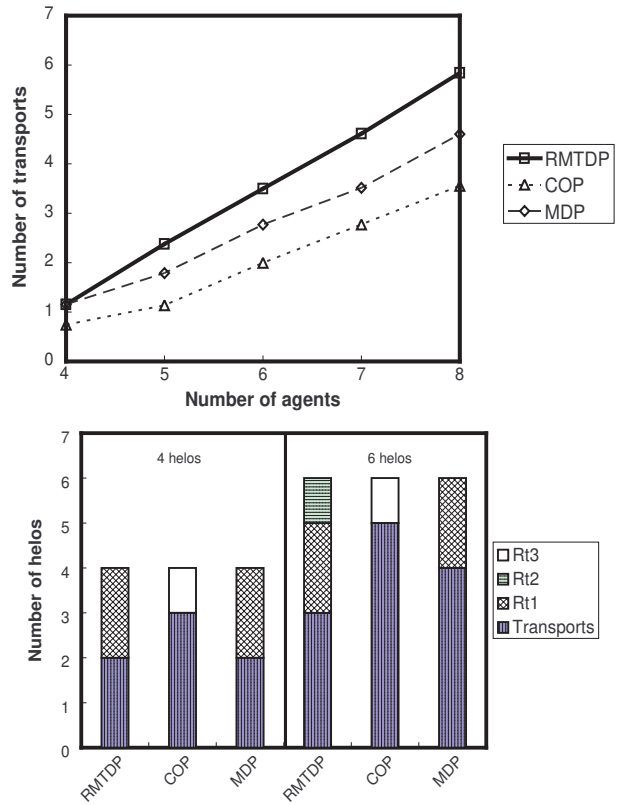

Figure 13: a) Comparison of performance of different allocation methods, b)Allocations found using different allocation methods.

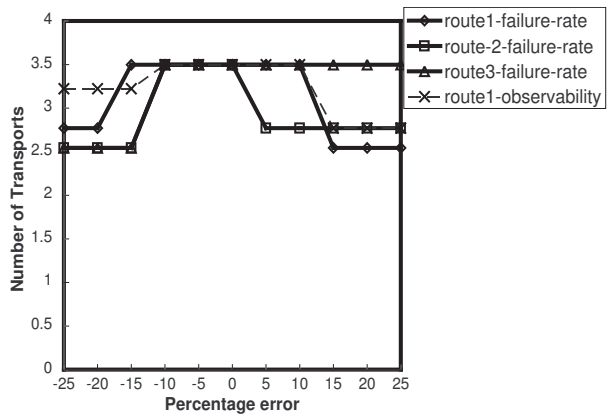

Figure 14: Model sensitivity in mission rehearsal domain.

## 6.2 Results in RoboCupRescue Domain

### 6.2.1 Speedups in RoboCupRescue Domain

In our next set of experiments, we highlight the computational savings obtained in the RoboCupRescue domain. The scenario for this experiment consisted of two fires at different locations in the city. Each of these fires has a different initially unknown number of civilians in it, however the total number of civilians and the distribution from which the locations of the civilians is chosen is known ahead of time. For this experiment, we fix the number of civilians at five and set the distribution used to choose the civilians' locations to be uniform. The number of fire engines is set at five, located in three different fire stations as described in Section 2.1 and vary the number of ambulances, all co-located at an ambulance center, from two to seven. The reason we chose to change only the number of ambulances is because small number of fire engines are unable to extinguish fires, changing the problem completely. The goal is to determine which fire engines to allocate to which fire and once information about civilians is transmitted, how many ambulances to send to each fire location.

Figure 15 highlights the savings in terms of the number of nodes evaluated and the actual runtime as we increase the number of agents. We show results only from NOPRUNE-BEL and MAXEXP. NOPRUNE-OBS could not be run because of slowness. Here the NOFAIL heuristic is identical to MAXEXP since agents cannot fail in this scenario. The RMTDP in this case had about 30,000 reachable states.

In both Figures 15(a) and 15(b), we increase the number of ambulances along the X-axis. In Figure 15(a), we show the number of nodes evaluated (parent nodes + leaf nodes)[5] on a logarithmic scale. As can be seen, the MAXEXP method results in about a 89-fold decrease in the number of nodes evaluated when compared to NOPRUNE-BEL for seven ambulances, and this decrease becomes more pronounced as the number of ambulances is increased. Figure 15(b) shows the time in seconds on a logarithmic scale on the Y-axis and compares the run-times of the MAXEXP and NOPRUNE-BEL methods for finding the best role allocation. The NOPRUNE-BEL method could not find the best allocation within a day when the number of ambulances was increased beyond four. For four ambulances (and five fire engines), MAXEXP resulted in about a 29-fold speedup over NOPRUNE-BEL.

### 6.2.2 Allocation in RoboCupRescue

Our next set of experiments shows the practical utility of our role allocation analysis in complex domains. We are able to show significant performance improvements in the actual RoboCupRescue domain using the role allocations generated by our analysis. First, we construct an RMTDP for the rescue scenario, described in Section 2.1, by taking guidance from the TOP and the underlying domain (as described in Section 4.1). We then use the MAXEXP heuristic to determine the best role allocation. We compared the RMTDP allocation with the allocations chosen by human subjects. Our goal in comparing RMTDP allocations with human subjects was mainly to show that RMTDP is capable at performing at or near human expert levels for this domain. In addition, in order to determine that reasoning about uncertainty actually impacts the allocations, we compared the RMTDP allocations with allocations determined by two additional allocation methods:

1. RescueISI: Allocations used by the our RoboCupRescue agents that were entered in the RoboCupRescue competitions of 2001(RescueISI) (Nair et al., 2002), where they finished in third place. These agents used local reasoning for their decision making, ignoring transitional as well and observational uncertainty.

2. RMTDP with complete observability: As discussed earlier, complete observability in RMTDP leads to an MDP, and we refer to this method as the MDP method.

---

5. The number of nodes evaluated using NOPRUNE-BEL can be computed as $(f_1 + 1) \cdot (f_2 + 1) \cdot (f_3 + 1) \cdot (a + 1)^{c+1}$, where $f_1$, $f_2$ and $f_3$ are the number of fire engines are station 1, 2 and 3, respectively, $a$ is the number of ambulances and c is the number of civilians. Each node provides a complete conditional role allocation, assuming different numbers of civilians at each fire station.

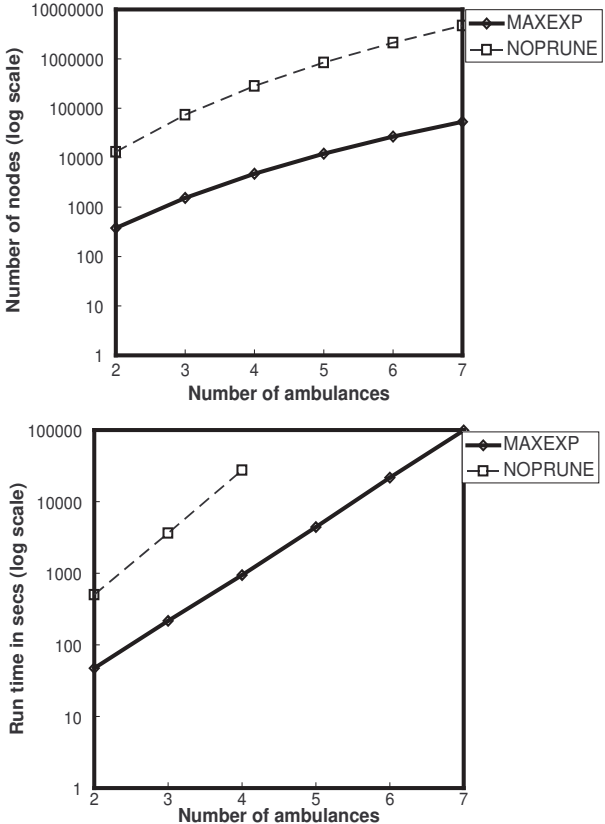

Figure 15: Performance of role allocation space search in RoboCupRescue, a: (left) Number of nodes evaluated on a log scale, and b: (right) Run-time in seconds on a log scale.

Note that these comparisons were performed using the RoboCupRescue simulator with multiple runs to deal with stochasticity[6]. The scenario is as described in Section 6.2.1. We fix the number of fire engines, ambulances and civilians at five each. For this experiment, we consider two settings, where the location of civilians is drawn from:

- Uniform distribution – 25% of the cases have four civilians at fire 1 and one civilian at fire 2, 25% with three civilians at fire 1 and two at fire 2, 25% with two civilians at fire 1 and three at fire 2 and the remaining 25% with one civilian at fire 1 and four civilians at fire 2. The speedup results of Section 6.2.1 were obtained using this distribution.

- Skewed distribution – 80% of the cases have four civilians at fire 1 and one civilian at fire 2 and the remaining 20% with one civilian at fire 1 and four civilians at fire 2.

Note that we do not consider the case where all civilians are located at the same fire as the optimal ambulance allocation is simply to assign all ambulances to the fire where the civilians are located. A skewed distribution was chosen to highlight the cases where it becomes difficult for humans to reason about what allocation to choose.

The three human subjects used in this experiment were researchers at USC. All three were familiar with RoboCupRescue. They were given time to study the setup and were not given any time limit to provide their allocations. Each subject was told that the allocations were going to be judged first on the basis of the number of civilian lives lost and next on the damage sustained due to fire. These are exactly the criteria used in RoboCupRescue (Kitano et al., 1999).

We then compared "RMTDP" allocation with those of the human subjects in the RoboCupRescue simulator and with RescueISI and MDP. In Figure 16, we compared the performance of the allocations on the basis of the number of civilians who died and the average damage to the two buildings (lower values are better for both criteria). These two criteria are the main two criteria used in RoboCupRescue (Kitano et al., 1999). The values shown in Figure 16 were obtained by averaging forty simulator runs for the uniform distribution and twenty runs for the skewed distribution for each allocation. The average values were plotted to account for the stochasticity in the domain. Error bars are provided to show the standard error for each allocation method.

As can be seen in Figure 16(a), the RMTDP allocation did better than the other five allocations in terms of a lower number of civilians dead (although human3 was quite close). For example, averaging forty runs, the RMTDP allocation resulted in 1.95 civilian deaths while human2's allocation resulted in 2.55 civilian deaths. In terms of the average building damage, the six allocations were almost indifferentiable, with the humans actually performing marginally better. Using the skewed distribution, the difference between the allocations was much more perceptible (see Figure 16(b)). In particular, we notice how the RMTDP allocation does much better than the humans in terms of the number of civilians dead. Here, human3 did particularly badly because of a bad allocation for fire engines. This resulted in more damage to the buildings and consequently to the number of civilians dead.

---

6. For the mission rehearsal domain, we could run on the actual mission rehearsal simulator since that simulator is not public domain and no longer accessible, and hence the difference in how we tested role allocations in the mission rehearsal and the RoboCupRescue domains.

Comparing RMTDP with RescueISI and the MDP approach showed that reasoning about transitional uncertainty (MDP) does better than a static reactive allocation method (RescueISI) but not as well as reasoning about both transitional and observational uncertainty. In the uniform distribution case, we found that RMTDP does better than both MDP and RescueISI, with the MDP method performing better than RescueISI. In the skewed distribution case, the improvement in allocations using RMTDP is greater. Averaging twenty simulation runs, RMTDP allocations resulted in 1.54 civilians deaths while MDP resulted in 1.98 and RescueISI in 3.52. The allocation method used by RescueISI often resulted in one of the fires being allocated too few fire engines. The allocations determined by the MDP approach turned out to be the same as human1.

A two-tailed t-test was performed in order to test the statistical significance of the means for the allocations in Figure 16. The means of number of civilians dead for the RMTDP allocation and the human allocations were found to be statistically different (confidence > 96%) for both the uniform as well as the skewed distributions. The difference in the fire damage was not statistically significant in the uniform case, however, the difference between the RMTDP allocation and human3 for fire damage was statistically significant (> 96%) in the skewed case.

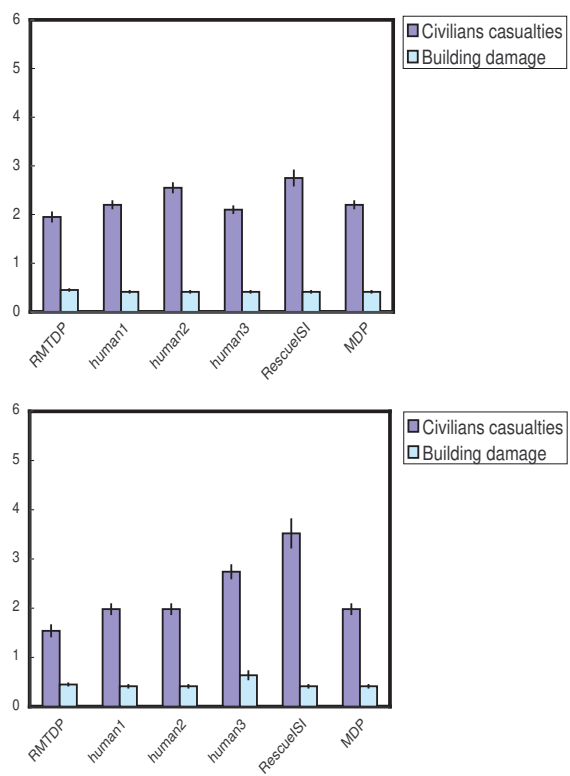

Figure 16: Comparison of performance in RoboCupRescue, a: (left) uniform, and b: (right) skewed.

Considering just the average performance of these different allocations does not highlight the individual cases where marked differences were seen in the performance. In Figure 17, we present the comparison of particular settings where the other allocation methods showed a bigger difference from RMTDP in terms of their allocations. The standard error is shown in error bars for each allocation. Figures 17(a) and 17(b) compare the allocations for uniform civilian distributions in the setting where there was one civilian at fire 1 and four civilians at fire 2 (1-4 civilian setting) and four civilians at fire 1 and one at fire 2 (4-1 civilian setting) respectively. As can be seen in these figure, the RMTDP allocation results in fewer civilian casualties but in slightly more damage to the buildings due to fire (difference in fire damage was not statistically significant because the damage values were very close). Figures 17(c) and 17(d) compare the allocations for the skewed civilian distribution. The key difference arises for human3. As can be seen, human3 results in more damage due to fire. This is because human3 allocated too few fire engines to one of the buildings, which in turn resulted in that building being burnt down completely. Consequently, civilians located at this fire location could not be rescued by the ambulances. Thus, we see specific instances where the allocation done using the RMTDP-based allocation algorithm is superior to allocations that a human comes up with.

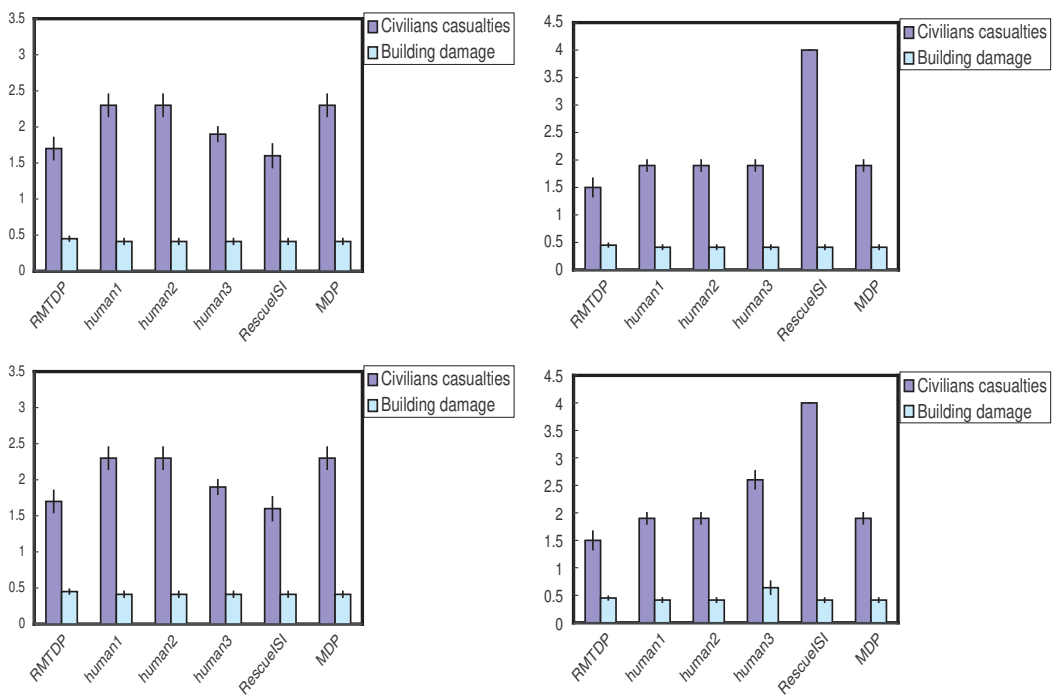

Figure 17: Comparison of performance in RoboCupRescue for particular settings, a: (top-left) uniform 1-4 civilian setting b:(top-right) uniform 4-1 civilian setting, c: (bottom-left) skewed 1-4 civilian setting d:(bottom-right) skewed 4-1 civilian setting.

Table 1 shows the allocations to fire 1 (agents not assigned to fire 1 are allocated to fire 2) found by the RMTDP role allocation algorithm and those used by the human subjects for the skewed 4-1 civilian setting (we consider this case since it shows the most difference). In particular, this table highlights the differences between the various allocators for the skewed 4-1 civilian setting and helps account for the differences seen in their performance in the actual simulator. As can be seen from Figure 17(d), the main difference in performance was in terms of the number of civilians saved. Recall that in this scenario, there are four civilians at fire 1, and one at fire 2. Here all the human subjects and MDP chose to send only one ambulance to fire 2 (*number of ambulances allocated to fire 2 = 5 − number of ambulances allocated to fire 1*). This lone ambulance was unable to rescue the civilian at fire 1, resulting in the humans and MDP saving fewer civilians. RescueISI chose to send all the ambulances to fire 2 using a greedy selection method based on proximity to the civilians resulting in all the civilians at fire 1 dying[7]. In terms of the fire engine allocation, human3 sent in four fire engines for fire 1 where more civilians were likely to be located (*number of engines allocated to fire 2 = 5 − number of engines allocated to fire 1*). Unfortunately, this backfired since the lone fire engine at fire 2 was not able to extinguish the fire there, causing the fire to spread to other parts of the city.

| Distribution | | RMTDP | human1 | human2 | human3 | RescueISI | MDP |
|---|---|---|---|---|---|---|---|
| Skewed 4-1 | Engines from station 1 | 0 | 2 | 2 | 1 | 2 | 2 |
| | Engines from station 2 | 1 | 1 | 1 | 1 | 1 | 1 |
| | Engines from station 3 | 1 | 0 | 0 | 2 | 0 | 0 |
| | Ambulances | 3 | 4 | 4 | 4 | 0 | 4 |

Table 1: Allocations of ambulances and fire engines to fire 1.

These experiments show that the allocations found by the RMTDP role allocation algorithm performs significantly better than allocations chosen by human subjects and RescueISI and MDP in most cases (and does not do significantly worse in any case). In particular when the distribution of civilians is not uniform, it is more difficult for humans to come up with an allocation and the difference between human allocations and the RMTDP allocation becomes more significant. From this we can conclude that the RMTDP allocation performs at near-human expertise.

In our last experiment done using the RoboCupRescue simulator, we introduced error in the RMTDP model in order to determine how sensitive the model was to errors in the parameters of the model. Figure 18 compares the allocations found, when there were five ambulances, 5 fire engines and 5 civilians, in terms of the number of civilian casualties (Y-axis) when error (X-axis) was introduced to the probability of fire spread and the probability of civilian health deterioration. As can be seen increasing the error in the probability of fire spread to 20% and higher results in allocations that save fewer civilians as the fire brigades choose to concentrate their effort on only one of the fires. The resulting allocation was found to have the same value in terms of the number of civilians casualties as that used by RescueISI, which did not consider any uncertainty. Reducing the error in the probability of fire did not have an impact on the allocations found. Increasing the error in probability of civilian health deterioration to 15% and higher caused some civilians to be sacrificed. This allocation was found to have the same value in terms of the number of civilians casualties as that used by RescueISI. Decreasing the error in probability of civilian health deterioration -5% and lower (more negative) caused the number of ambulances to be allocated to a fire to be the same as the number of civilians at that fire (same as human1).

---

7. This strategy of ambulances going to the closest civilian worked fairly well because the ambulances were usually well spread out

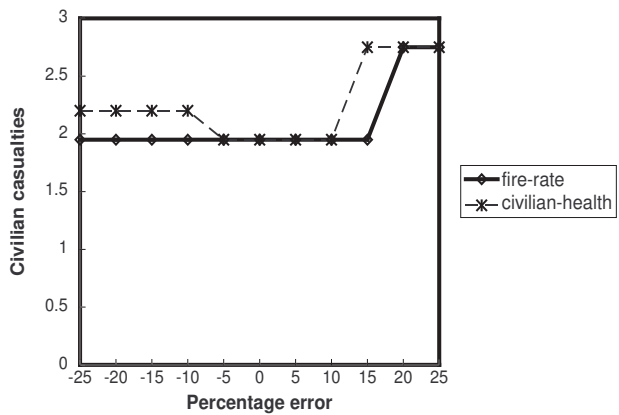

Figure 18: Model sensitivity in the RoboCupRescue scenario.

# 7. Related Work

There are four related areas of research, that we wish to highlight. First, there has been a considerable amount of work done in the field of multiagent teamwork (Section 7.1). The second related area of research is the use of decision theoretic models, in particular distributed POMDPs (Section 7.2). The third area of related work we describe (Section 7.3) are hybrid systems that used Markov Decision Process and BDI approaches. Finally, in Section 7.4, the related work in role allocation and reallocation in multiagent teams is described.

## 7.1 BDI-based Teamwork

Several formal teamwork theories such as *Joint Intentions* (Cohen & Levesque, 1991), *SharedPlans* (Grosz & Kraus, 1996) were proposed that tried to capture the essence of multiagent teamwork in the logic of Beliefs-Desires-Intentions. Next, practical models of teamwork such as COLLAGEN (Rich & Sidner, 1997), GRATE* (Jennings, 1995), STEAM (Tambe, 1997) built on these teamwork theories (Cohen & Levesque, 1991; Grosz & Kraus, 1996) and attempted to capture the aspects of teamwork that were reusable across domains. In addition, to complement the practical teamwork models, the team-oriented programming approach (Pynadath & Tambe, 2003; Tidhar, 1993a, 1993b) was introduced to allow large number of agents to be programmed as teams. This approach was then expanded on and applied to a variety of domains (Pynadath & Tambe, 2003; Yen et al., 2001; da Silva & Demazeau, 2002). Other approaches for building practical multia-gent systems (Stone & Veloso, 1999; Decker & Lesser, 1993), while not explicitly based on team-oriented programming, could be considered in the same family.

The research reported in this article complements this research on teamwork by introducing hybrid BDI-POMDP models that exploit the synergy between BDI and POMDP approaches. In particular, TOP and teamwork models have traditionally not addressed uncertainty and cost. Our hybrid model provides this capability, and we have illustrated the benefits of this reasoning via detailed experiments.

While this article uses team-oriented programming (Tambe et al., 2000; da Silva & Demazeau, 2002; Tidhar, 1993a, 1993b) as an example BDI approach, it is relevant to other similar techniques of modeling and tasking collectives of agents, such as Decker and Lesser's (1993) TAEMS approach. In particular, the TAEMS language provides an abstraction for tasking collaborative groups of agents similar to TOP, while the GPGP infrastructure used in executing TAEMS-based tasks is analogous to the "TOP interpreter" infrastructure shown in Figure 1. While Lesser *et al.* have explored the use of distributed MDPs in analyses of GPGP coordination (Xuan & Lesser, 2002), they have not exploited the use of TAEMS structures in decomposition or abstraction for searching optimal policies in distributed MDPs, as suggested in this article. Thus, this article complements Lesser *et al.*'s work in illustrating a significant avenue for further efficiency improvements in such analyses.

## 7.2 Distributed POMDP Models

Distributed POMDP models represent a collection of formal models that are expressive enough to capture the uncertainty in the domain and the costs and rewards associated with states and actions. Given a group of agents, the problem of deriving separate policies for them that maximize some joint reward can be modeled using distributed POMDP models. In particular, the DEC-POMDP (**Dec**entralized POMDP) (Bernstein et al., 2000) and MTDP (**M**ultiagent **T**eam **D**ecision **P**roblem) (Pynadath & Tambe, 2002) are generalizations of POMDPs to the case where there are multiple, distributed agents, basing their actions on their separate observations. These frameworks allow us to formulate what constitutes an optimal policy for a multiagent team and in principle derive that policy.

However, with a few exceptions, effective algorithms for deriving policies for distributed POMDPs have not been developed. Significant progress has been achieved in efficient single-agent POMDP policy generation algorithms (Monahan, 1982; Cassandra, Littman, & Zhang, 1997; Kaelbling et al., 1998). However, it is unlikely such research can be directly carried over to the distributed case. Finding optimal policies for distributed POMDPs is NEXP-complete (Bernstein et al., 2000). In contrast, finding an optimal policy for a single agent POMDP is PSPACE-complete (Papadimitriou & Tsitsiklis, 1987). As Bernstein *et al.* note (Bernstein et al., 2000), this suggests a fundamental difference in the nature of the problems. The distributed problem cannot be treated as one of separate POMDPs in which individual policies can be generated for individual agents because of possible cross-agent interactions in the reward, transition or observation functions. (For any one action of one agent, there may be many different rewards possible, based on the actions that other agents may take.)

Three approaches have been used to solve distributed POMDPs. One approach that is typically taken is to make simplifying assumptions about the domain. For instance, in Guestrin *et al.* (2002), it is assumed that each agent can completely observe the world state. In addition, it is assumed that the reward function (and transition function) for the team can be expressed as the sum (product) of the reward (transition) functions of the agents in the team. Becker *et al.* (2003) assume that the domain is factored such that each agent has a completely observable local state and also that the domain is transition-independent (one agent cannot affect another agent's local state).

The second approach taken is to simplify the nature of the policies considered for each of the agents. For example, Chadès et al. (2002) restrict the agent policies to be memoryless (reactive) policies, thereby simplifying the problem to solving multiple MDPs. Peshkin *et al.* (2000) take a different approach by using gradient descent search to find local optimum finite-controllers with bounded memory. Nair *et al.* (2003a) present an algorithm for finding a locally optimal policy from a space of unrestricted finite-horizon policies. The third approach, taken by Hansen *et al.* (2004), involves trying to determine the globally optimal solution without making any simplifying assumptions about the domain. In this approach, they attempt to prune the space of possible complete policies by eliminating dominated policies. Although a brave frontal assault on the problem, this method is expected to face significant difficulties in scaling up due to the fundamental complexity of obtaining a globally optimal solution.

The key difference with our work is that our research is focused on hybrid systems where we leverage the advantages of BDI team plans, which are used in practical systems, and distributed POMDPs that quantitatively reason about uncertainty and cost. In particular, we use TOPs to specify large-scale team plans in complex domains and use RMTDPs for finding the best role allocation for these teams.

### 7.3 Hybrid BDI-POMDP Approaches

POMDP models have been used in the context of analysis of both single agent (Schut, Wooldridge, & Parsons, 2001) and multiagent (Pynadath & Tambe, 2002; Xuan et al., 2001) behavior. Schut *et al.* compare various strategies for intention reconsideration (deciding when to deliberate about its intentions) by modeling a BDI system using a POMDP. The key differences with this work and our approach are that they apply their analysis to a single agent case and do not consider the issues of exploiting BDI system structure in improving POMDP efficiency.

Xuan and Lesser (2001) and Pynadath and Tambe (2002), both analyze multiagent communication. While Xuan and Lesser dealt with finding and evaluating various communication policies, Pynadath and Tambe used the COM-MTDP model to deal with the problem of comparing various communication strategies both empirically and analytically. Our approach is more general in that we explain an approach for analyzing any coordination actions including communication. We concretely demonstrate our approach for analysis of role allocation. Additional key differences from the earlier work by Pynadath and Tambe (2002) are as follows: (i) In RMTDP, we illustrate techniques to exploit team plan decomposition in speeding up policy search, absent in COM-MTDP, (ii) We also introduce techniques for belief-based evaluation absent from previous work. Nonetheless, combining RMTDP with COM-MTDP is an interesting avenue for further research and some preliminary steps in this direction are presented in Nair, Tambe and Marsella (2003b).

Among other hybrid systems not focused on analysis, Scerri *et al.* (2002) employ Markov Decision Processes within team-oriented programs for adjustable autonomy. The key difference between that work and ours is that the MDPs were used to execute a particular sub-plan within the TOP's plan hierarchy and not for making improvements to the TOP. DTGolog (Boutilier, Reiter, Soutchanski, & Thrun, 2000) provides a first-order language that limits MDP policy search via logical constraints on actions. Although it shares with our work the key idea of synergistic interactions in MDPs and Golog, it differs from our work in that it focuses on single agent MDPs in fully observable domains, and does not exploit plan structure in improving MDP performance. ISAAC (Nair, Tambe, Marsella, & Raines, 2004), a system for analyzing multiagent teams, also employs decision theoretic methods for analyzing multiagent teams. In that work, a probabilistic finite automaton (PFA) that represents the probability distribution of key patterns in the team's behavior are learned from logs of the team's behaviors. The key difference with that work is that the analysis is performed without having access to the actual team plans that the agents are executing and hence the advice provided cannot directly be applied to improving the team, but will need a human developer to change the team behavior as per the advice generated.

## 7.4 Role Allocation and Reallocation

There are several different approaches to the problem of role allocation and reallocation. For example, Tidhar *et al.* (1996) and Tambe *et al.* (2000) performed role allocation based on matching of capabilities, while Hunsberger and Grosz (2000) proposed the use of combinatorial auctions to decide on how roles should be assigned. Modi *et al.* (2003) showed how role allocation can be modeled as a distributed constraint optimization problem and applied it to the problem of tracking multiple moving targets using distributed sensors. Shehory and Kraus (1998) suggested the use of coalition formation algorithms for deciding quickly which agent took on which role. Fatima and Wooldridge (2001) use auctions to decide on task allocation. It is important to note that these competing techniques are not free of the problem of how to model the problem, even though they do not have to model transition probabilities. Other approaches to reforming a team are reconfiguration methods due to Dunin-Keplicz and Verbrugge (2001), self-adapting organizations by Horling and Lesser (2001) and dynamic re-organizing groups (Barber & Martin, 2001). Scerri *et al.* (2003) present a role (re)allocation algorithm that allows autonomy of role reallocation to shift between a human supervisor and the agents.

The key difference with all this prior work is our use of stochastic models (RMTDPs) to evaluate allocations: this enables us to compute the benefits of role allocation, taking into account uncertainty and costs of reallocation upon failure. For example, in the mission rehearsal domain, if uncertainties were not considered, just one scout would have been allocated, leading to costly future reallocations or even in mission failure. Instead, with lookahead, depending on the probability of failure, multiple scouts were sent out on one or more routes, resulting in fewer future reallocations and higher expected reward.

## 8. Conclusion

While the BDI approach to agent teamwork has provided successful applications, tools and techniques that provide quantitative analyses of team coordination and other team behaviors under uncertainty are lacking. The emerging field of distributed POMDPs provides a decision theoretic method for quantitatively obtaining the optimal policy for a team of agents, but faces a serious intractability challenge. Therefore, this article leverages the benefits of both the BDI and POMDP approaches to analyze and improve key coordination decisions within BDI-based team plans using POMDP-based methods. In order to demonstrate these analysis methods, we concentrated on role allocation – a fundamental aspect of agent teamwork – and provided three key contributions. First, we introduced RMTDP, a distributed POMDP based framework, for analysis of role allocation. Second, this article presented an RMTDP-based methodology for optimizing key coordination decisions within a BDI team plan for a given domain. Concretely, the article described a methodology for finding the best role allocation for a fixed team plan. Given the combinatorially many role allocations, we introduced methods to exploit task decompositions among sub-teams to significantly prune the search space of role allocations.

Third, our hybrid BDI-POMDP approach uncovered several synergistic interactions between BDI team plans and distributed POMDPs:

1. TOPs were useful in constructing the RMTDP model for the domain, in identifying the features that need to be modeled as well as in decomposing the model construction according to the structure of the TOP. The RMTDP model could then be used to evaluate the TOP.

2. TOPs restricted the policy search by providing RMTDPs with incomplete policies with a limited number of open decisions.

3. The BDI approach helped in coming up with a novel efficient "belief-based" representation of policies suited for this hybrid BDI-POMDP approach and a corresponding algorithm for evaluating such policies. This resulted in faster evaluation and also a more compact policy representation.

4. The structure in the TOP was exploited to decompose the problem of evaluating abstract policies, resulting in significant pruning in the search for the optimal role allocations.

We constructed RMTDPs for two domains – RoboCupRescue and mission rehearsal simulation – and determined the best role allocation in these domains. Furthermore, we illustrated significant speedups in RMTDP policy search due to the techniques introduced in this article. Detailed experiments revealed the advantages of our approach over state-of-the-art role allocation approaches that failed to reason with uncertainty.

Our key agenda for future work is to continue scale-up of RMTDPs to even larger scale agent teams. Such scale-up will require further efficiency improvements. We propose to continue to exploit the interaction in the BDI and POMDP approaches in achieving such scale-up. For instance, besides disaster rescue, distributed sensor nets and large area monitoring applications could benefit from such a scale-up.

## Acknowledgments

This research was supported by NSF grant #0208580. We would like to thank Jim Blythe, Anthony Cassandra, Hyuckchul Jung, Spiros Kapetanakis, Sven Koenig, Michael Littman, Stacy Marsella, David Pynadath and Paul Scerri for discussions related to this article. We would also like to thank the reviewers of this article whose comments have helped in significantly improving this article.

## Appendix A. TOP details

In this section, we will describe the TOP for the helicopter scenario. The details of each subplan in Figure 4(b) are shown below:

```
ExecuteMission:
Context:∅
Pre-conditions:  (MB <TaskForce> location(TaskForce) = START)
Achieved:  (MB <TaskForce> (Achieved(DoScouting) ∧ Achieved(DoTransport)))
           ∧ (time > T ∨ (MB <TaskForce> Achieved(RemainingScouts) ∨
           (∄ helo ∈ ScoutingTeam, alive(helo) ∧ location(helo) ≠ END)))
Unachievable:  (MB <TaskForce> Unachievable(DoScouting))
               ∨ (MB <TaskForce> (Unachievable(DoTransport)
               ∧ (Achieved(RemainingScouts)
               ∨(∄ helo ∈ ScoutingTeam, alive(helo) ∧ location(helo) ≠ END))))
Irrelevant:  ∅
Body:
DoScouting
DoTransport
RemainingScouts
Constraints:
DoScouting → DoTransport
DoScouting → RemainingScouts

DoScouting:
Context:ExecuteMission <TaskForce>
Pre-conditions:  ∅
Achieved:  ∅
Unachievable:  ∅
Irrelevant:∅
Body:
WaitAtBase
ScoutRoutes
Constraints:
WaitAtBase AND ScoutRoutes

WaitAtBase:
Context:  DoScouting <TaskForce>
Pre-conditions:  ∅
Achieved:  ∅
Unachievable:  (MB <TransportTeam> ∄ helo ∈ TransportTeam, alive(helo))
```

411

```
Irrelevant:  ∅
Body:
no-op

ScoutRoutes:
Context:  DoScouting <TaskForce>
Achieved:  ∅
Unachievable:  ∅
Irrelevant:(MB <ScoutingTeam> ∄ helo ∈ TransportTeam, alive(helo))
Body:
ScoutRoute1
ScoutRoute2
ScoutRoute3
Constraints:
ScoutRoute1 OR ScoutRoute2 OR ScoutRoute3

ScoutRoute1:
Context:  ScoutRoutes <ScoutingTeam>
Pre-conditions:  ∅
Achieved:  (MB <SctTeamA> ∃ helo ∈ SctTeamA, location(helo) = END)
Unachievable:  time > T ∨ (MB <SctTeamA> ∄ helo ∈ SctTeamA, alive(helo))
Irrelevant:  ∅
Body:
if (location(SctTeamA) = START) then route(SctTeamA) ← 1
if (location(SctTeamA) ≠ END) then move-forward

ScoutRoute2:
Context:  ScoutRoutes <ScoutingTeam>
Pre-conditions:  ∅
Achieved:  (MB <SctTeamB> ∃ helo ∈ SctTeamB, location(helo) = END)
Unachievable:  time > T ∨ (MB <SctTeamB> ∄ helo ∈ SctTeamB, alive(helo))
Irrelevant:  ∅
Body:
if (location(SctTeamB) = START) then route(SctTeamB) ← 2
if (location(SctTeamB) ≠ END) then move-forward

ScoutRoute2:
Context:  ScoutRoutes <ScoutingTeam>
Pre-conditions:  ∅
Achieved:  (MB <SctTeamA> ∃ helo ∈ SctTeamA, location(helo) = END)
Unachievable:  time > T ∨ (MB <SctTeamA> ∄ helo ∈ SctTeamA, alive(helo))
Irrelevant:  ∅
Body:
if (location(SctTeamA) = START) then route(SctTeamA) ← 1
if (location(SctTeamA) ≠ END) then move-forward

DoTransport:
Context:  ExecuteMission <TaskForce>
Pre-conditions:  ∅
```

```
Achieved:  (MB <TransportTeam> location(TransportTeam) = END)
Unachievable:  time > T ∨ (MB <TransportTeam> ∄ helo ∈ TransportTeam, alive(helo))
Irrelevant:  ∅
Body:
if (location(TransportTeam) = start) then
  if (MB <TransportTeam> Achieved(ScoutRoute1)) then
    route(TransportTeam) ← 1
  elseif (MB <TransportTeam> Achieved(ScoutRoute2)) then
    route(TransportTeam) ← 2
  elseif (MB <TransportTeam> Achieved(ScoutRoute3)) then
    route(TransportTeam) ← 3
if (route(TransportTeam) ≠ null) and (location(TransportTeam) ≠ END) then
    move-forward

RemainingScouts:
Context:  ExecuteMission <TaskForce>
Pre-conditions:  ∅
Achieved:  (MB <ScoutingTeam> location(ScoutingTeam) = END)
Unachievable:  time > T ∨ (MB <ScoutingTeam> (∄ helo ∈ ScoutingTeam
              alive(helo) ∧ location(helo) ≠ END))
Irrelevant:  ∅
Body:
if (location(ScoutingTeam) ≠ END) then move-forward
```

The predicate Achieved(tplan) is true if the Achieved conditions of *tplan* are true. Similarly, the predicates Unachievable(tplan) and Irrelevant(tplan) are true if the the Unachievable conditions and the Irrelevant conditions of *tplan* are true, respectively. The predicate (location(team) = END) is true if all members of *team* are at END.

Figure 4(b) also shows coordination relationships: An AND relationship is indicated with a solid arc, while an OR relationship is indicated with a dotted arc. These coordination relationships indicate unachievability, achievability and irrelevance conditions that are enforced by the TOP infrastructure. An AND relationship between team sub-plans means that if any of the team sub-plans fail, then the parent team plan will fail. Also, for the parent team plan to be achieved, all the child sub-plans must be achieved. Thus, for **DoScouting**, **WaitAtBase** and **ScoutRoutes** must both be done:

```
Achieved:  (MB <TaskForce> Achieved(WaitAtBase) ∧ Achieved(ScoutRoutes))
Unachievable:  (MB <TaskForce> Unachievable(WaitAtBase)
              ∨ Unachievable(ScoutRoutes))
```

An OR relationship means that all the subplans must fail for the parent to fail and success of any of the subplans means that the parent plan has succeeded. Thus, for **ScoutingRoutes**, at least one of **ScoutRoute1**, **ScoutRoute2** or **ScoutRoute3** need be performed:

```
Achieved:  (MB <ScoutingTeam> Achieved(ScoutRoute1) ∨
          Achieved(ScoutRoute2)∨ Achieved(ScoutRoute3))
Unachievable:  (MB <TaskForce> Unachievable(ScoutRoute1) ∧
              Unachievable(ScoutRoute2) ∧ Unachievable(ScoutRoute3))
```

Also an AND relationship affects the irrelevance conditions of the subplans that it joins. If the parent is unachievable then all its subplans that are still executing become irrelevant. Thus, for **WaitAtBase**:

```
Irrelevant:  (MB <TaskForce> Unachievable(ScoutRoutes))
```

Similarly for **ScoutingRoutes**:

```
Irrelevant:  (MB <TaskForce> Unachievable(ScoutRoutes))
```

.

Finally, we assign roles to plans — Figure 4(b) shows the assignment in brackets adjacent to the plans. For instance, *Task Force* team is assigned to jointly perform **Execute Mission**.

## Appendix B. RMTDP details

In this section, we present details of the RMTDP constructed for the TOP in Figure 4.

- $S$: We get the features of the state from the attributes tested in the preconditions and achieved, unachievable and irrelevant conditions and the body of the team plans and individual agent plans. Thus the relevant state variables are:location of each helicopter, role of each helicopter,route of each helicopter, status of each helicopter (alive or not) and time. For a team of $n$ helicopters, the state is given by the tuple $< time, role_1, \ldots, role_n, loc_1, \ldots, loc_n, route_1, \ldots, route_n, status_1, \ldots, status_n >$.

- $A$: We consider actions to be the primitive actions that each agent can perform within its individual plans. The TOP infrastructure enforces mutual belief through communication actions. Since analyzing the cost of these is not the focus of this research we consider communication to be implicit and we model the effect of this communication directly in the observation function.

  We consider 2 kinds of actions role-taking and role-execution actions. We assume that the initial allocation will specify roles for all agents. This specifies whether the agent is a scout or a transport and if a scout which scout team it is assigned to. A scout cannot become a transport or change its team after its initial allocation while a transport can change its role by taking one of the role-taking actions.The role-taking and role-execution actions for each agent $i$ are given by:
  $\Upsilon_{i,memberTransportTeam} = \{joinSctTeamA, joinSctTeamB, joinSctTeamC\}$
  $\Upsilon_{i,memberSctTeamA} = \Upsilon_{i,memberSctTeamB} = \Upsilon_{i,memberSctTeamCx} = \emptyset$
  $\Phi_{i,memberTransportTeam} = \{chooseRoute, moveForward\}$
  $\Phi_{i,memberSctTeamA} = \Phi_{i,memberSctTeamB} = \Phi_{i,memberSctTeamC} = \{moveForward\}$

- $P$: We obtain the transition function with the help of a human expert or through simulations if a simulator is available. In this domain, helicopters can crash (be shot down) if they are not at START, END or an already scouted location. The probability that scouts will get shot down depends on which route they are on, i.e. probability of crash on route1 is $p_1$, probability of crash on route2 is $p_2$ and probability of crash on route3 is $p_3$ and how many scouts are on the same spot. We assume that the probability of a transport being shot down in an unscouted location to be 1 and in a scouted location to be 0. The probability of multiple crashes can be obtained by multiplying the probabilities of individual crashes.

The action, moveForward, will have no effect if $route_i = \textbf{null}$ or $loc_i = \textbf{END}$ or if $status_i = \textbf{dead}$. In all other cases, the location of the agent gets incremented. We assume that the role-taking actions $scoutRoutex$ will always succeed if the role of the performing agent is $transport$ and it has not been assigned a route already.

- $\Omega$: Each transport at START can observe the status of the other agents with some probability depending on their positions. Each helicopter on a particular route can observe all the helicopters on that route completely and cannot observe helicopters on other routes.

- $O$: The observation function gives the probability for a group of agents to receive a particular joint observation. In this domain we assume that observations of one agent are independent of the observations of other agents, given the current state and the previous joint action. Thus the probability of a joint observation can be computed by multiplying the probabilities of each individual agent's observations.

  The probability of a transport at START observing the status of an alive scout on route 1 is 0.95. The probability of a transport at START observing nothing about that alive scout is 0.05 since we don't have false negatives. Similarly if a scout on route 1 crashes, the probability that this is visible to a transport at START is 0.98 and the probability that the transport doesn't see this failure is 0.02. Similarly the probabilities for observing an alive scout on route 2 and route 3 and 0.94 and 0.93 respectively and the probabilities for observing a crash on route 2 and route 3 and 0.97 and 0.96 respectively.

- $R$: The reward function is obtained with the help of a human expert who helps assign value to the various states and the cost of performing various actions. For this analysis, we assume that actions moveForward and chooseRoute have no cost. We consider the negative reward (cost) for the replacement action, scoutRoute$x$, to be $R_\Upsilon$, the negative reward for a failure of a helicopter to be $R_F$, the reward for a scout reaching END to be $R_{scout}$ and the reward for a transport reaching END to be $R_{transport}$. E.g. $R_\Upsilon = -10$, $R_F = -50$, $R_{scout} = 5$, $R_{transport} = 75$.

- $\mathcal{RL}$: These are the roles that individual agents can take in TOP organization hierarchy. $\mathcal{RL} = \{transport, scoutOnRoute1, scoutOnRoute2, scoutOnRoute3\}$.

## Appendix C. Theorems

**Theorem 3** *The MAXEXP method will always yield an upper bound.*

**Proof sketch:**

- Let policy $\pi^*$ be the leaf-level policy with the highest expected reward under a particular parent node, $i$, in the restricted policy space.

$$V_{\pi^*} = \textbf{max}_{\pi \in Children(i)} V_\pi \tag{3}$$

- Since the reward function is specified separately for each component, we can separate the expected reward $V$ into the rewards from the constituent components given the starting states and starting observation histories of these components. Let the team plan be divided into $m$ components such that the components are parallel and independent or sequentially executed.

$$V_{\pi^*} \leq \sum_{1 \leq j \leq m} \mathbf{max}_{states[j],oHistories[j]} V_{j\pi^*}$$

- The expected value obtained for any component $j$, $1 \leq j \leq m$ for $\pi^*$ cannot be greater than that of the highest value obtained for $j$ using any policy.

$$\mathbf{max}_{states[j],oHistories[j]} V_{j\pi^*} \leq \mathbf{max}_{\pi \in Children(i)} \mathbf{max}_{states[j],oHistories[j]}(V_{j\pi}) \qquad (4)$$

- Hence,

$$V_{\pi^*} \leq \sum_{1 \leq j \leq m} \mathbf{max}_{\pi \in Children(i)} \mathbf{max}_{states[j],oHistories[j]}(V_{j\pi})$$

$$V_{\pi^*} \leq \mathbf{MaxEstimate}(i) \qquad (5)$$

$\square$